\documentclass[11pt,letterpaper]{article}
\pdfoutput=1

\usepackage{jcappub}
\usepackage{amsmath}
\usepackage{amsfonts}
\usepackage{amssymb}
\usepackage{graphicx}
\usepackage{epsfig}
\usepackage{multirow}

\def\be{\begin{equation}}
\def\ee{\end{equation}}
\def\bea{\begin{eqnarray}}
\def\eea{\end{eqnarray}}
\def\ba{\begin{eqnarray}}
\def\ea{\end{eqnarray}}

\title{Constraining the initial conditions of the Universe using large scale structure}

\author[a]{Nishant Agarwal,}
\author[a]{Shirley Ho,}
\author[b]{and Sarah Shandera}
\affiliation[a]{McWilliams Center for Cosmology, Department of Physics, Carnegie Mellon University, \\ Pittsburgh, PA 15213, USA}
\affiliation[b]{Institute for Gravitation and the Cosmos, Pennsylvania State University, \\ University Park, PA 16802, USA}

\emailAdd{nishanta@andrew.cmu.edu}
\emailAdd{shirleyh@andrew.cmu.edu}
\emailAdd{shandera@gravity.psu.edu}

\abstract{Primordial non-Gaussianity induces a scale-dependent bias in large scale structure (LSS) data, proportional to $f_{\rm NL}/k^2$ for the exact local ansatz. Recent work has shown that models of inflation that predict a large squeezed limit bispectrum, such as multi-field models and single field inflation with a modified initial state, typically give rise to a generalized local ansatz, with the scale-dependent bias now proportional to ${\cal A}_{\rm NL}/k^\alpha$. We use photometric measurements of the angular power spectrum of luminous red galaxies and quasars in the Sloan Digital Sky Survey Data Release Eight (SDSS DR8) with the above parameterization to constrain the amplitude ${\cal A}_{\rm NL}$ and scale-dependence $\alpha$. We find that the marginalized upper limit on $\alpha$ is 2.0 at the $95\%$ confidence level, consistent with the local ansatz. We also present Fisher forecasts for a survey of the same size as DR8 to assess the role of systematics in current photometric LSS data. Moreover, we present analytic results on the expected mass dependence of ${\cal A}_{\rm NL}$ for different inflationary models, which can be an important observable for future surveys, if primordial non-Gaussianity is non-zero.}

\preprint{IGC-13/10-1}

\begin{document}

\maketitle


\section{Introduction}
\label{sec:intro}

The spectrum of primordial curvature perturbations (denoted with $\zeta$) generated during inflation is very nearly Gaussian. Current data from the cosmic microwave background (CMB) and large scale structure (LSS), however, is detailed enough to measure even small deviations from Gaussianity. Interactions of the inflaton and other primordial fields leave very distinct non-Gaussian features in statistics of the temperature anisotropies in the CMB and the density fluctuations that collapse into bound objects.

The power spectrum $P_{\zeta}(k)$ is defined in terms of the two-point function in momentum space,
\bea
	\big\langle \zeta_{\vec{k}_{1}} \zeta_{\vec{k}_{2}} \big\rangle & = & (2\pi)^{3} \delta^{3} \big( \vec{k}_{1} + \vec{k}_{2} \big) P_{\zeta} (k_{1}),
\eea
and has been well-measured on large scales using the CMB. In an isotropic Universe, the lowest order statistic that can reveal primordial `non-Gaussianity' is the three-point function of the perturbations, or the bispectrum. Analogous to the power spectrum, the bispectrum $B_{\zeta} (k_{1}, k_{2}, k_{3})$ is defined using the three-point function,
\bea
	\big\langle \zeta_{\vec{k}_{1}} \zeta_{\vec{k}_{2}} \zeta_{\vec{k}_{3}} \big\rangle & = & (2\pi)^{3} \delta^{3} \big( \vec{k}_{1} + \vec{k}_{2} + \vec{k}_{3} \big) B_{\zeta} (k_{1}, k_{2}, k_{3}).
\eea
Homogeneity, enforced by the Dirac delta functions in the expressions above, requires that the momenta in the bispectrum form a triangle. The assumption of isotropy further ensures that the power spectrum and the bispectrum will depend only on magnitudes of the momenta. Different inflationary models for generating the primordial fluctuations can be identified by the `shape' of the resulting bispectrum, defined as the type of triangle for which the amplitude of the bispectrum is largest. So far the best method to detect primordial non-Gaussianity has been to measure the three-point function of temperature anisotropies in the CMB (see \cite{Ade:2013ydc} for recent CMB constraints on $f_{\rm NL}$ from the Planck satellite and \cite{Hinshaw:2012aka,Bennett:2012zja} for CMB constraints using nine-year data from the Wilkinson Microwave Anisotropy Probe (WMAP) satellite). While the most recent bounds have tightened the limits on non-Gaussianity by a significant factor, the data has certainly not yet excluded any complete class of inflationary scenarios. Non-Gaussianity remains a key observable for understanding primordial physics, and the very large galaxy surveys planned over the next decade provide an opportunity to push constraints past the theoretically interesting thresholds.

As with the CMB temperature fluctuations, the statistics of matter density fluctuations in LSS are sensitive to primordial non-Gaussianity. We expect the LSS three-point function to eventually provide powerful constraints, but we so far lack a precise understanding of how to extract primordial quantities from the bispectrum of late time luminous objects. However, it has been realized that one need not measure the complete bispectrum to uncover important physical information: the region of parameter space where one of the momenta is much smaller than the others (the so-called {\it squeezed} limit, $k_{3} \ll k_{1} \approx k_{2}$) is both a good diagnostic of the theory {\it and} a simpler limit to constrain with LSS data. It was observed in \cite{Dalal:2007cu} (also see \cite{Grinstein:1986en,Matarrese:1986et} for earlier work) that non-Gaussianity with significant coupling in the squeezed limit affects the abundance and clustering of virialized objects, so that another (less obvious) observable is the {\it power spectrum} of dark matter halos (see \cite{Ross:2012sx,Karagiannis:2013xea,Ho:2013lda} and section 3 of this paper for recent LSS constraints on $f_{\rm NL}$ using data from the Sloan Digital Sky Survey (SDSS) \cite{York:2000gk,Eisenstein:2011sa}; also see \cite{Giannantonio:2013uqa} for constraints from CMB-galaxy cross correlations). Dark matter halos are associated with peaks in the initial, linear density field, whose heights exceed some threshold. In the peak-background split formalism \cite{Cole:1989vx} the density field is separated into long and short wavelength modes, and the large scale clustering of halos is estimated by computing the effect of long wavelength modes on small scales. In the presence of primordial non-Gaussianities, large and small scale density fluctuations are not necessarily independent, and this mode coupling leads to an additional scale-dependent term in the halo bias. Any primordial bispectrum leads to a shift in the halo bias whose precise form and observability depends on how the small scale power in fluctuations is correlated with long wavelength fluctuations.

The (Eulerian) bias, $b(M,k,z,f_{\rm NL})$, with $M$ the mass of the halo that the LSS tracer follows, is defined as the ratio of the matter-halo cross power spectrum to the matter power spectrum,
\bea
	b(M,k,z,f_{\rm NL}) & \equiv & \frac{P_{\rm matter-halo}(M,k,z,f_{\rm NL})}{P_{\rm matter}(k,z)} \\
	& = & b_{1}(M,z) + \Delta b_{{\rm non-Gaussian}}.
\eea
Here the scale of observation, $k$, corresponds to the long mode in the squeezed limit (i.e. $k \equiv k_{3}$). With purely Gaussian density fluctuations, the bias of a halo of mass $M$ at redshift $z$ approaches the scale-independent value, $b_{1}(M,z)$. 

A very useful example of a non-Gaussian scenario that gives rise to a significant signal in the halo bias is the well-studied `local ansatz', where the gravitational potential is a simple non-linear function of the {\it local} value of a Gaussian field. The strength of the correlation between modes of different wavelengths is parameterized by a constant $f_{\rm NL}$,
\bea
	\Phi_{\vec{k}} &  = & \Phi_{\vec{k},{\rm Gaussian}} + f_{\rm NL} \left( \Phi_{\vec{k},{\rm Gaussian}}^{2} - \left\langle \Phi_{\vec{k},{\rm Gaussian}}^{2} \right\rangle \right),
\eea
or equivalently,
\bea
	\zeta_{\vec{k}} &  = & \zeta_{\vec{k},{\rm Gaussian}} + \frac{3}{5} f_{\rm NL} \left( \zeta_{\vec{k},{\rm Gaussian}}^{2} - \left\langle \zeta_{\vec{k},{\rm Gaussian}}^{2} \right\rangle \right).
\eea
Here $\Phi_{\vec{k}}$ is (minus) the primordial gravitational potential (in the matter-dominated era) and the subscript indicates a Gaussian random field. The above parameterization generates a bispectrum of the form \cite{Gangui:1993tt,Verde:1999ij,Komatsu:2001rj},
\bea
	B_{\zeta,{\rm local}} (k_{1}, k_{2}, k_{3}) & = & \frac{6}{5} f_{\rm NL} \left[ P_{\zeta}(k_{1}) P_{\zeta}(k_{2}) + 2 \ {\rm perm.} \right],
\label{eq:bispectrum}
\eea
which has a maximum amplitude in the squeezed limit. For the local form bispectrum it was found that the scale-dependent correction to the bias is given by \cite{Dalal:2007cu,Matarrese:2008nc,Slosar:2008hx}
\bea
	\Delta b(M,k,z,f_{\rm NL}) & = & 3f_{\rm NL}[b_{1}(M,z) - p]\delta_{c} \frac{\Omega_{m}H_{0}^{2}}{k^{2}T(k)D(z)}.
\label{eq:deltabstd}
\eea
Here, $\delta_{c} \approx 1.686$ denotes the critical density for spherical collapse, $\Omega_{m}$ is the present day matter density, $H_{0}$ is the Hubble constant, $T(k)$ is the matter transfer function normalized to unity as $k \rightarrow 0$, and $D(z)$ is the linear growth function normalized to $(1+z)^{-1}$ in the matter dominated era. The parameter $p$ ranges from unity for objects that populate all halos equally to $1.6$ for objects that populate only recently merged halos \cite{Slosar:2008hx}. The predicted scale-dependent term in the halo bias has been used to constrain local $f_{\rm NL}$ using data from different tracers of LSS, such as luminous red galaxies (LRGs) and quasars \cite{Slosar:2008hx, Xia:2010yu,Xia:2010pe,Xia:2011hj,Ross:2012sx,Karagiannis:2013xea,Ho:2013lda}. It has recently been suggested \cite{Tashiro:2012wr} that 21 cm observations and observations of the Lyman$-\alpha$ forest may also give interesting constraints on primordial non-Gaussianity in the near future.

For arbitrary non-Gaussian scenarios, the leading term in the non-Gaussian halo bias has the more general form \cite{Shandera:2010ei,Desjacques:2011jb,Desjacques:2011mq}
\bea
	\Delta b \left( M,k,z,{\mathcal A}_{\rm NL},\alpha \right) \propto [b_{1}(M,z) - p] \frac{{\mathcal A}_{\rm NL} \left( b_{1}(M,z) \right)}{k^{\alpha}}.
\label{eq:genNGbias}
\eea
The power $\alpha$ can be read off from the dominant dependence of {\it any} bispectrum on the long wavelength mode ($k_{3}$). The dependence of the amplitude ${\mathcal A}_{\rm NL}$ on mass, or the observable $b_{1}$, on the other hand, comes from any dependence of the strength of the correlation on scale or additional dependence on short scale power. There are generically additional terms in $\Delta b$ that are subdominant in the long wavelength limit. While they should be included in testing any particular scenario, here we focus on the two-parameter ansatz for the leading term.

The non-Gaussian bias is important in distinguishing between different models of inflation. Standard single-field slow-roll inflation cannot produce observable non-Gaussianity in squeezed triangles \cite{Maldacena:2002vr,Creminelli:2004yq,Creminelli:2011rh,Pajer:2013ana}, so any detection of a non-Gaussian halo bias would rule out those scenarios. Inflationary scenarios involving multiple scalar fields can correlate long and short wavelength modes. Examples of models involving multiple fields that can enhance local non-Gaussianity include two field and $N$-field inflation \cite{Linde:1996gt,Bernardeau:2002jy,Seery:2005gb,Rigopoulos:2005us,Vernizzi:2006ve,Battefeld:2006sz,
Yokoyama:2007uu,Yokoyama:2007dw,Sasaki:2008uc,Byrnes:2008wi,Byrnes:2010em,Kim:2010ud,
Peterson:2010mv,Elliston:2011et, Dias:2012nf}, the curvaton scenario \cite{Lyth:2002my,Ichikawa:2008iq,Beltran:2008aa,Chambers:2009ki,Byrnes:2009pe,Enqvist:2009ww,
Alabidi:2010ba}, and inhomogeneous reheating \cite{Dvali:2003em,Zaldarriaga:2003my,Suyama:2007bg}. The strength of the coupling and the scale-dependence $\alpha$ depend on how the fields interact and evolve during inflation, and on the masses of the fields involved. The halo bias expected from generalizations of the local ansatz related to multi-field inflation models was studied in \cite{Desjacques:2009jb,Shandera:2010ei,Smith:2011ub,LoVerde:2011iz,Dias:2013rla}. Current multi-field models can generate $0\leq\alpha\lesssim2\pm\mathcal{O}(\epsilon)$ where $\epsilon<1$ is the slow-roll parameter. When all fields involved in the dynamical generation of the perturbations are light compared to the inflationary Hubble scale ($m\ll H$), $\alpha$ is closer to 2. When at least one field is light but others are heavy ($m\lesssim H$), $\alpha$ is closer to 0 \cite{Chen:2009we,Chen:2009zp}. In addition, single field models with modifications to the initial quantum state can generate non-Gaussianity of the squeezed type, at least over some range of scales \cite{Chen:2006nt,Holman:2007na,Chen:2008wn,Meerburg:2009ys,Agullo:2010ws,Ashoorioon:2010xg,Ganc:2011dy,Kundu:2011sg,
Chialva:2011hc,Agarwal:2012mq,Flauger:2013hra,Aravind:2013lra,Ashoorioon:2013eia,Kundu:2013gha}. In this case it appears that even $\alpha=3$ is allowed \cite{Ganc:2012ae,Agullo:2012cs}, although the most generic or natural bispectra in those cases are not yet fully understood. Previous constraints and forecasts based on LSS and CMB data, for various subsets of these models, can be found in \cite{Shandera:2010ei,Becker:2010hx,Sefusatti:2012ye,Norena:2012yi,Becker:2012yr,Becker:2012je,Giannantonio:2013uqa}.

Any non-Gaussianity that can be detected using the halo bias also has interesting ramifications for how we use observations to constrain inflation theory \cite{Nelson:2012sb,Nurmi:2013xv,LoVerde:2013xka,Byrnes:2013qjy}. Properties of the primordial fluctuations within our observed volume of the Universe may themselves be biased with respect to the mean statistics predicted by any inflationary model. Since we have no way of knowing whether or not our observed Universe has typical, mean, or highly biased statistics, local type non-Gaussianity introduces a new source of cosmic variance uncertainty in relating observations to theory. Tighter observational constraints, especially on small scales \cite{Bramante:2013moa}, are required to eliminate this cosmic variance as relevant for our cosmology. 

Since inflation model building continues and there is so far no hard theory limit on the range of $\alpha$ allowed by any conceivable model of inflation, we adopt the parameterization of eq.\ (\ref{eq:genNGbias}) to study observational constraints on ${\mathcal A}_{\rm NL}$ and $\alpha$ using data from LRGs and quasars in the SDSS-III Data Release Eight (DR8) sample \cite{Aihara:2011sj,Ross:2011cz,Ho:2012vy,Ho:2013lda}. We use these constraints to infer what current LSS observations tell us about the initial conditions in the very early Universe, and inflationary mechanisms for generating them. We also perform a Fisher matrix analysis to assess how much better could a survey of a similar volume as DR8 do, in the absence of any systematic uncertainties in the data.

With current LSS data, especially in the absence of a complete understanding of various systematics in the data, it appears difficult to probe the mass-dependence of the amplitude of non-Gaussianity. As a pointer for future surveys, we present analytic results on the expected scaling of the amplitude ${\cal A}_{\rm NL}$ with bias $(b_{1} - p)$, for different forms of the bispectrum.

The paper is organized as follows. In section \ref{sec:methoddata} we describe the data that we use and the method that we adopt. We obtain constraints on the scale-dependence of the bias in section \ref{sec:results}. We present Fisher forecasts in section \ref{sec:fisher} and in section \ref{sec:theory} we discuss the analytic method to determine the non-Gaussian correction to the bias and use it to obtain the scaling of ${\cal A}_{\rm NL}$ with $b_{1} - p$. We conclude with a discussion in section \ref{sec:discussion}.


\section{Method and data}
\label{sec:methoddata}

We begin by describing the general method we adopt to constrain models of inflation using LSS data. We would like to use a generalized version of eq.\ (\ref{eq:deltabstd}) given by
\bea
	\Delta b \left( M,k,z,{\mathcal A}_{\rm NL},\alpha \right) & = & 3{\mathcal A}_{\rm NL}(b_{1}(M,z)) [b_{1}(M,z) - p] \frac{\Omega_{m}H_{0}^{2}}{k^{2}(k/k_{p})^{\alpha-2}T(k)D(z)},
\label{eq:deltab}
\eea
in the halo power spectrum,
\bea
	P_{\rm halo} \left( M,k,z,{\mathcal A}_{\rm NL},\alpha \right) & = & \left[ b_{1}(M,z) + \Delta b \left( M,k,z,{\mathcal A}_{\rm NL},\alpha \right) \right]^{2} P_{\rm matter}(k,z), \quad
\label{eq:pkhalo}
\eea
and fit the corresponding angular power spectrum (rather than the full three-dimensional power spectrum above which is relatively difficult to measure) to LSS data, to constrain ${\mathcal A}_{\rm NL}$ and $\alpha$. For this purpose, we use a Markov-Chain Monte-Carlo (MCMC) approach to explore the available parameter space using a modified version of the widely used package {\tt CosmoMC} \cite{Lewis:2002ah}. We calculate the linear matter power spectrum using the {\tt CAMB} code \cite{Lewis:1999bs} included in the {\tt CosmoMC} package, and apply the HaloFit prescription \cite{Smith:2002dz} to account for non-linear effects on the matter power spectrum. In eq.\ (\ref{eq:deltab}) we have included $\delta_{c}$ in our definition of ${\mathcal A}_{\rm NL}$ to allow for deviations from the spherical collapse model. The pivot scale $k_{p}$ is chosen to be $0.1 \ {\rm Mpc}^{-1}$. This choice is based on the smoothing scale (defined in section \ref{sec:theory}) that approximately corresponds to a Gaussian bias of $b_{1} = 2.0$ (we use the Sheth-Tormen mass function \cite{ST1,ST2} to relate the variance at any smoothing mass scale to the Gaussian bias). In section \ref{sec:fisher} we will also use a Fisher matrix analysis to find the pivot points where ${\mathcal A}_{\rm NL}$ and $\alpha$ would be uncorrelated for optimally clean SDSS data.

To calculate the theoretical angular power spectrum, we use the full Bessel integration on the largest scales and account for redshift space distortions as described, for example, in \cite{Padmanabhan:2006ku},
\bea
\label{eq:Cell}
	C_{\ell} & = & C_{\ell}^{\rm gg} + C_{\ell}^{\rm gv} + C_{\ell}^{\rm vv} + a.
\eea
The superscripts ${\rm g}$ and ${\rm v}$ denote galaxy and velocity terms, respectively, and $a$ is an extra (constant) shot noise-like term that we add to obtain a better fit to the non-linear power spectrum \cite{dePutter:2012sh}. The three contributions to the angular power spectrum above are given by the integrals \cite{Padmanabhan:2006ku}
\bea
	C_{\ell}^{\rm gg} & = & \frac{2}{\pi} \int {\rm d}\ln k \ k^{3} P_{\rm matter}(k,0) W_{\ell}^{2}(k), \\
	C_{\ell}^{\rm gv} & = & \frac{4}{\pi} \int {\rm d}\ln k \ k^{3} P_{\rm matter}(k,0) W_{\ell}(k) W_{\ell}^{r}(k), \\
	C_{\ell}^{\rm vv} & = & \frac{2}{\pi} \int {\rm d}\ln k \ k^{3} P_{\rm matter}(k,0) \left[ W_{\ell}^{r}(k) \right]^{2},
\eea
with the window functions here calculated using
\bea
	W_{\ell}(k) & = & \int {\rm d}z \ (b_{1} + \Delta b) \frac{D(z)}{D(0)} \frac{{\rm d}N}{{\rm d}z} j_{\ell}(kr), \\
	W_{\ell}^{r}(k) & = & \int {\rm d}z \ \Omega_{m}^{0.56}(z) \frac{D(z)}{D(0)} \frac{{\rm d}N}{{\rm d}z} \Bigg[ \frac{2\ell^{2}+2\ell-1}{(2\ell-1)(2\ell+3)} j_{\ell}(kr) \nonumber \\
	& & \quad \quad - \ \frac{\ell(\ell-1)}{(2\ell-1)(2\ell+1)}j_{\ell-2}(kr) - \frac{(\ell+1)(\ell+2)}{(2\ell+1)(2\ell+3)}j_{\ell+2}(kr) \Bigg],
\eea
where we have suppressed the functional dependence of $b_{1}(M,z)$, $\Delta b \left( M,k,z,{\mathcal A}_{\rm NL},\alpha \right)$, and $r(z)$ for brevity of notation. Here ${\rm d}N/{\rm d}z$ is the redshift distribution normalized to unity, $r(z)$ is the comoving distance, and $j_{\ell}(kr)$ is the $\ell^{\rm th}$ order spherical Bessel function. In the flat sky (large $\ell$) limit, we switch to the Limber approximation \cite{Limber:1954zz}.

The theoretical spectrum obtained above is used to calculate the likelihood (assumed Gaussian), that is the input to the MCMC procedure,
\bea
	\chi^{2} & = & ({\bf d} - {\bf t})^{T} \ . \ {\mathcal C}^{-1} \ . \ ({\bf d} - {\bf t}).
\eea
Here ${\bf d}$ is the data $C_{\ell}$ vector, ${\bf t}$ is the theory $C_{\ell}$ vector convolved with the full survey window function, and ${\mathcal C}$ is the covariance matrix. The data vector here is calculated using an optimal quadratic estimator \cite{Seljak:1997ep,Tegmark:1997yq,Padmanabhan:2002yv,Padmanabhan:2006ku} which, although designed to compute nearly anti-correlated power spectra across different multipole bins, does retain a very small contribution ($\lesssim 5\%$) from other multipole bins. This makes it especially important to convolve the theoretical spectrum with the full window function (before calculating $\chi^{2}$) in non-zero ${\cal A}_{\rm NL}$ models, since the power spectrum rises dramatically at low $\ell$ in these models.

Maximizing the likelihood in the full parameter space in our MCMC analysis provides constraints on ${\mathcal A}_{\rm NL}$ and $\alpha$. We always use standard cosmological data, including the WMAP nine-year CMB data \cite{Hinshaw:2012aka,Bennett:2012zja} and the ``Union 2'' supernova data set that includes 557 supernovae \cite{Amanullah:2010vv}, as our baseline model. These data sets are not directly sensitive to the levels of non-Gaussianity considered in this paper; however, they are needed to constrain the basic cosmological model and thus the shape and normalization of the matter power spectrum.

We now turn to the LSS data used in this paper, that includes observations of LRGs and quasars in SDSS-III DR8 \cite{Aihara:2011sj,Ross:2011cz,Ho:2012vy,Ho:2013lda}. The SDSS has mapped over a quarter of the sky using the dedicated Sloan Foundation $2.5$ m optical telescope \cite{Gunn:2006tw} located at the Apache Point Observatory in New Mexico, USA. A drift-scanning mosaic CCD camera \cite{Gunn:1998vh} images the sky in five bands ($ugriz$) \cite{Fukugita:1996qt,Smith:2002pca} to a limiting magnitude of around $22.5$. The data is then processed through a series of pipelines that perform astrometric calibration \cite{Pier:2002iq}, photometric reduction \cite{Lupton:2001zb}, and photometric calibration \cite{Padmanabhan:2007zd}. For our analysis, we only use photometric angular power spectra of LRGs and quasars in SDSS-III DR8, which we describe below. The underlying redshift distributions, however, are calculated using the Baryon Oscillation Spectroscopic Survey (BOSS) \cite{Dawson:2012va,Bolton:2012hz,Smee:2012wd} spectroscopic redshifts of the same sample.

\subsection{Photometric LRGs from SDSS}
\label{subsec:datalrgs}

The LRG data and the method for obtaining angular power spectra are described in \cite{Ross:2011cz,Ho:2012vy}. We refer the reader to these papers for details and provide only a brief description of the main properties here.

The data set spans $\sim 11,000$ square degrees of the sky and probes a volume of $\sim 3h^{-3} \ {\rm Gpc^{3}}$. We focus on the approximately stellar mass-limited CMASS sample of luminous galaxies, that follows the CMASS galaxy selection detailed in \cite{White:2010ed}. Photometric redshifts and the probability that an object is a galaxy are obtained using a training sample of 112,778 BOSS CMASS spectra. The final catalog consists of 872,921 luminous galaxies in the redshift range $0.45 \leq z \leq 0.65$. The estimated photometric error increases from 0.04 to 0.06 over the redshift range. Four photometric bins are defined with widths similar to the error --- $z = 0.45-0.50$, $0.50-0.55$, $0.55-0.60$, and $0.60-0.65$, with the effective number of galaxies (weighting each object with the probability that it is a galaxy) in each bin being $214971$, $258736$, $248895$, and $150319$ respectively. The normalized redshift distribution in these four bins can be found in \cite{Ross:2011cz,Ho:2012vy}.

The calculation of the angular power spectra in the four redshift bins uses the optimal quadratic estimator method outlined in \cite{Seljak:1997ep,Tegmark:1997yq,Padmanabhan:2002yv,Padmanabhan:2006ku}. The four power spectra are binned in $\ell$ space with a typical wave band width of $\Delta \ell = 10$. We plot the power spectra with their error bars in the next section (fig.\ \ref{fig1}).

We first determine the Gaussian bias in each redshift slice, for use later in the paper. We only use those $\ell$ bins in the LRG angular power spectra that have a contamination of less than $3\sigma$ from unknown systematics, as defined in terms of the cross-power between different redshift slices in \cite{Agarwal:2013ajb}.\footnote{The method presented in \cite{Agarwal:2013ajb} is a generalization of that discussed in \cite{Ross:2011cz,Ho:2012vy}, to include the effects of both known and unknown systematics. For other ways of dealing with systematics see, e.g., \cite{Huterer:2012zs,Pullen:2012rd,Giannantonio:2013uqa,Hernandez-Monteagudo:2013vwa,Leistedt:2013gfa}.} We also choose a low-$\ell$ cutoff at $\ell_{\rm min} = 10$ as we expect lower multipoles to be dominated with systematics, and a high-$\ell$ cutoff, $\ell_{\rm max}$, corresponding to $k = 0.1 h \ {\rm Mpc}^{-1}$ (determined using $\Lambda{\rm CDM}$ cosmology, see table \ref{table1}) to avoid the strongly non-linear regime of the matter power spectrum. An MCMC analysis over the standard cosmological parameters $\left\{ \Omega_{b}h^{2}, \Omega_{\rm DM}h^{2}, \theta, \tau, n_{s}, \log A_{s}, A_{\rm SZ} \right\}$, the bias for each redshift slice, and the four corresponding non-linear fitting parameters, with ${\cal A}_{\rm NL} = 0$, using WMAP9 + SN + DR8 (LRG) data, yields the best-fit bias values shown in table\ \ref{table1}. Here $\Omega_{b}h^{2}$ is the physical baryon density, $\Omega_{\rm DM}h^{2}$ is the physical dark matter density, $\theta$ is the ratio of the sound horizon to the angular diameter distance at decoupling, $\tau$ is the reionization optical depth, $n_{s}$ is the scalar spectral index, $A_{s}$ is the amplitude of the primordial scalar curvature perturbations at $k =0.05 \ {\rm Mpc}^{-1}$, and $A_{\rm SZ}$ represents a Sunyaev-Zeldovich template normalization.

\begin{table}[!h]
\begin{center}
	\begin{tabular}{|c|c|c|c|}
		\hline
		$z_{\rm mid}$ & $l_{\rm max}$ & ${\rm Gaussian \ bias}, b_{1}$ & $10^{6}a$ \\
		& $(k = 0.1 h \ {\rm Mpc}^{-1})$ & & \\
		\hline
		0.475 & 128 & $1.96^{+0.14}_{-0.14}$ & $2.86^{+5.78}_{-5.79}$ \\
		0.525 & 140 & $1.97^{+0.11}_{-0.10}$ & $3.83^{+3.32}_{-3.30}$ \\
		0.575 & 151 & $2.09^{+0.12}_{-0.13}$ & $2.14^{+3.10}_{-3.08}$ \\
		0.625 & 162 & $2.26^{+0.11}_{-0.12}$ & $1.44^{+2.35}_{-2.35}$ \\
		\hline
	\end{tabular}
\caption{The best-fit Gaussian bias and the non-linear fitting parameter in eq.\ (\ref{eq:Cell}) (with $1\sigma$ errors) in the four redshift slices for LRGs, using WMAP9 + SN + DR8 (LRG) data. Here we set ${\cal A}_{\rm NL} = 0$ and use only those $\ell$ bins in $10 \le \ell \le \ell_{\rm max}$ that satisfy a $3\sigma$-cut on unknown systematics \cite{Agarwal:2013ajb}.}
\label{table1}
\end{center}
\end{table}

\subsection{Photometric quasars from SDSS}
\label{subsec:dataqsos}

We use photometric quasars from \cite{Ho:2013lda}, and refer the reader to this paper for details. Here we summarize some of the main properties of the data.

The data set spans $\sim 11,000$ square degrees of the sky and traces a volume of $\sim 80h^{-3} \ {\rm Gpc^{3}}$, larger than the volume probed by LRGs since quasars, being amongst the most luminous objects in the Universe, can be observed to much higher redshifts. A total of 822 BOSS spectra are used to estimate the true redshifts for the photometric catalog. The final catalog is based on a sample of $409,914$ quasars over the redshift range $0.5 \leq z \leq 2.5$.  The data is divided into four redshift bins --- $z = 0.5-1.0$, $1.0-1.5$, $1.5-2.0$, and $2.0-2.5$, with the effective number of quasars in each bin being $47710$, $142096$, $148166$, and $71942$ respectively. The normalized redshift distribution in these four bins is given in \cite{Ho:2013lda}.

The calculation of the quasar angular power spectra in the four redshift bins uses the optimal quadratic estimator as well, and the power spectra are binned in $\ell$ space with a typical width of $\Delta \ell = 20$. We plot the power spectra with their error bars in the next section (fig.\ \ref{fig2}).

As for LRGs, we first determine the Gaussian bias for quasars in the four redshift slices, for use later in the paper. We use only those $\ell$ bins in the quasar angular power spectra that have a contamination of less than $1\sigma$ from unknown systematics, as defined in \cite{Agarwal:2013ajb,Ho:2013lda} and choose a low-$\ell$ cutoff at $\ell_{\rm min} = 10$ and a high-$\ell$ cutoff, $\ell_{\rm max}$, corresponding to $k = 0.1 h \ {\rm Mpc}^{-1}$ (determined using $\Lambda{\rm CDM}$ cosmology, see table \ref{table2}). In the MCMC analysis we vary over the standard cosmological parameters and the bias in each redshift slice, while setting the non-linear fitting parameter in each redshift slice to zero as the error bars on the data are too large to allow for a good fit to this parameter. Using WMAP9 + SN + DR8 (quasar) data we find the best-fit bias values shown in table\ \ref{table2}.

Many bins in the first redshift slice appear to be dominated by unknown systematics and are subsequently dropped. Using the remaining bins gives an estimate of the bias (which is varied in $b_{1} \in [0.1,10]$) that is not bounded from below (see table\ \ref{table2}). In further analysis we therefore use the {\it median} value of the bias for the first redshift slice, which is $b_{1} = 2.57$ instead of the {\it mean} value of $b_{1} = 2.19$.

\begin{table}[!h]
\begin{center}
	\begin{tabular}{|c|c|c|}
		\hline
		$z_{\rm mid}$ & $l_{\rm max}$ & ${\rm Gaussian \ bias}, b_{1}$ \\
		& $(k = 0.1 h \ {\rm Mpc}^{-1})$ & \\
		\hline
		0.75 & 189 & $2.19^{+0.47}_{-2.09}$ \\
		1.25 & 278 & $2.06^{+0.08}_{-0.08}$ \\
		1.75 & 346 & $2.32^{+0.11}_{-0.09}$ \\
		2.25 & 400 & $3.37^{+0.20}_{-0.18}$ \\
		\hline
	\end{tabular}
\caption{The best-fit Gaussian bias (with $1\sigma$ errors) in the four redshift slices for quasars, using WMAP9 + SN + DR8 (quasar) data. Here we set ${\cal A}_{\rm NL} = 0$ and use only those $\ell$ bins in $10 \le \ell \le \ell_{\rm max}$ that satisfy a $1\sigma$-cut on unknown systematics \cite{Agarwal:2013ajb,Ho:2013lda}.}
\label{table2}
\end{center}
\end{table}


\section{Results}
\label{sec:results}

In this section we use the LSS data described in sections \ref{subsec:datalrgs} and \ref{subsec:dataqsos} to constrain ${\cal A}_{\rm NL}$ and $\alpha$. We first need to fix the parameter $p$ in eq.\ (\ref{eq:deltab}). The halo occupation distribution (HOD) for LRGs depends only on the mass of the halo, so we set $p=1$ for them. Quasar activity, on the other hand, is believed to be triggered by recent mergers. The HOD for quasars, therefore, depends not only on the final mass of the halo but also on the formation history. In the extreme case that quasars only populate recently merged halos, it was found in \cite{Slosar:2008hx} that $p=1.6$. Their analysis was generalized in \cite{Reid:2010vc} to include a dependence on the halo formation redshift. Here we ignore these effects and set $p=1.6$ for quasars in our analysis.

The LRG constraints in this section are based only on those $\ell$ bins in $10 \le \ell \le \ell_{\rm max}$ that satisfy a $3\sigma$-cut on unknown contamination and the quasar constraints are based on the bins that satisfy a $1\sigma$-cut, as in section \ref{sec:methoddata}. Further, we set the bias to the best-fit values from tables\ \ref{table1} and \ref{table2} and fix the $a$ parameter in each redshift slice for LRGs to the best-fit value in the Gaussian case from table\ \ref{table1}.

In principle, all free parameters should be varied in the MCMC analysis. However, for Gaussian likelihoods it is reasonable to set the bias and $a$ parameters to their best-fit values, which also offers significant speed advantages. Since we will be varying over both the amplitude and the scale-dependence of the non-Gaussian bias this assumption is not strictly valid. Nevertheless, it is a good starting point since we do not expect large deviations in the Gaussian bias, which was fit using a wide range of multipoles, while the scale-dependent correction has a characteristic shape that depends on the value of $\alpha$. In section \ref{subsec:constraintsfNL} we perform a simple test to check the validity of this assumption for $\alpha = 2$.

\subsection{Constraints on $f_{\rm NL}$}
\label{subsec:constraintsfNL}

We begin by reviewing the constraints for the exact local ansatz, $\alpha = 2$, in which case ${\cal A}_{\rm NL}/\delta_{c}$ simply reduces to local $f_{\rm NL}$. These results were presented in \cite{Agarwal:2013ajb,Ho:2013lda}. On performing an MCMC analysis over standard cosmological parameters and $f_{\rm NL}$ using different data sets, we obtain the constraints shown in table\ \ref{table3}.

\begin{table}[!h]
\begin{center}
	\begin{tabular}{|l|l|}
		\hline
		\multicolumn{1}{|c|}{Data set} & \multicolumn{1}{|c|}{$f_{\rm NL}$} \\
		\hline
		LRGs & $-17^{+68}_{-68}$ \\
		Quasars & $103^{+148}_{-146}$ \\
		LRGs + quasars & $2^{+65}_{-66}$ \\
		\hline
	\end{tabular}
\caption{Constraints on $f_{\rm NL}$ for the exact local ansatz ($\alpha = 2$), with $68\%$ confidence limits, using different data sets.}
\label{table3}
\end{center}
\end{table}

Owing to the large volume traced by quasars, we would expect tight constraints on primordial non-Gaussianity from quasar angular power spectra. However, we find that much of the data at low $\ell$ is contaminated by unknown systematics. Since we drop all $\ell$ bins with significant systematic uncertainties in the cross-correlations, the constraints from quasars are not very strong. For the same reason these constraints are weaker compared to earlier analyses that made use of SDSS quasars, such as \cite{Slosar:2008hx,Xia:2011hj,Xia:2010pe}.

\begin{figure}[!h]
\begin{center}
	\includegraphics[width=6.0in,angle=0]{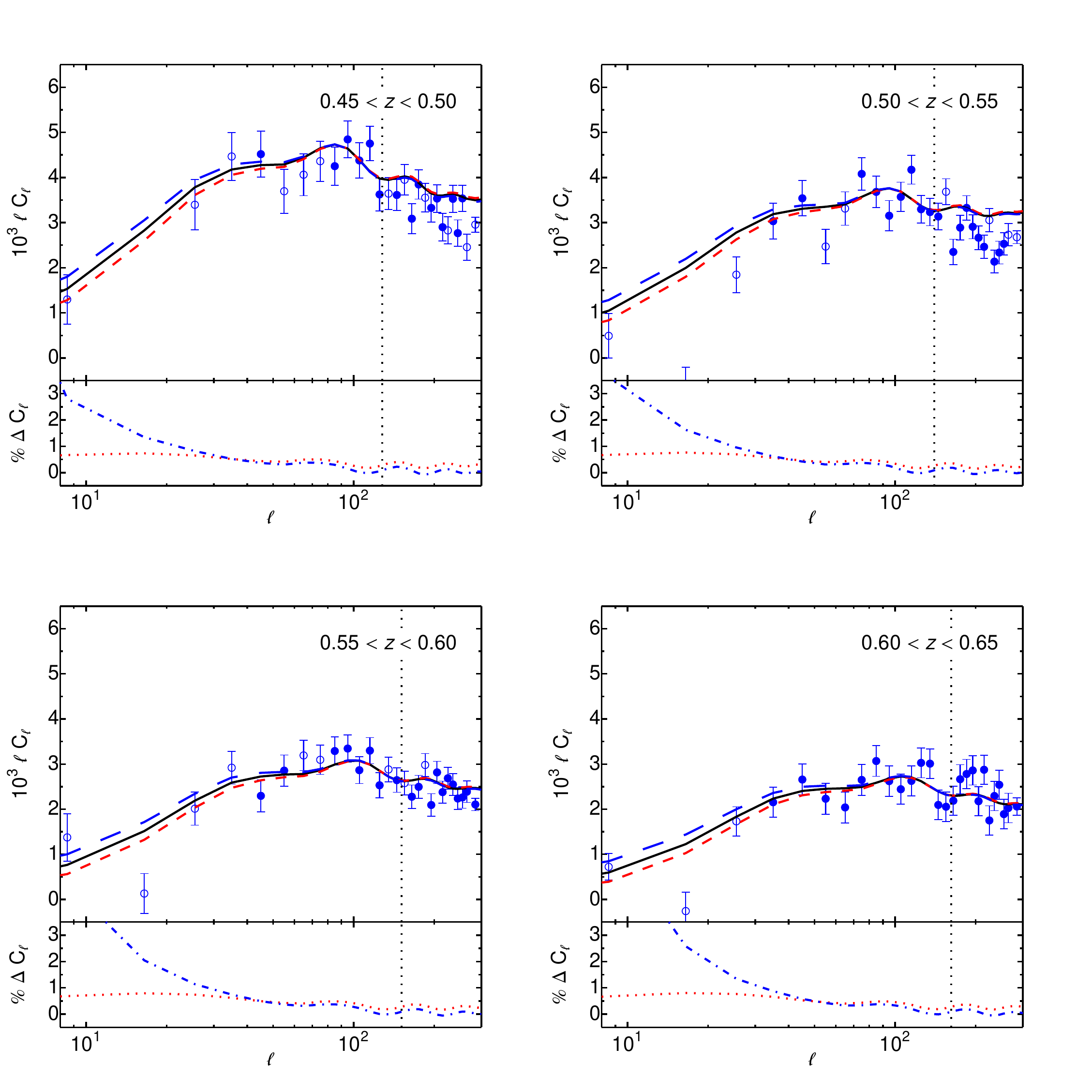}
	\caption{The angular power spectrum in the four redshift slices of LRGs. Open circles with ($1\sigma$) error bars represent data points that are excluded due to large unknown systematics, as determined using the cross-power spectra. Filled circles, on the other hand, are data points that are not dominated with unknown systematics. Note that in each redshift slice there are some bins that would not appear contaminated but are still dropped as their cross-power with another redshift slice is significantly contaminated and one cannot tell a priori which redshift slice is responsible for the contamination. The vertical dotted line shows $\ell_{\rm max}$; we only use filled data points in $10 \leq \ell \leq \ell_{\rm max}$ for our analysis. The curves in the upper panels are the theoretical angular power spectra at the best-fit and $68\%$ confidence values of $f_{\rm NL}$ for LRGs + quasars --- $f_{\rm NL} = 2$ (solid black), $f_{\rm NL} = -64$ (dashed red), and $f_{\rm NL} = 67$ (long-dashed blue). We also add the non-linear fitting parameter $a$ to the theoretical $C_{\ell}$s, which causes the upturn at large $\ell$. The curves in the lower panels are the percentage differences of the best-fit angular power spectrum for LRGs + quasars with respect to the Gaussian angular power spectrum --- ${\cal A}_{\rm NL} = -1$ with $\alpha = 1.7$ (dotted red) and ${\cal A}_{\rm NL} = 0.9$ with $\alpha = 3$ (dot-dashed blue).}
\label{fig1}
\end{center}
\end{figure}

\begin{figure}[!h]
\begin{center}
	\includegraphics[width=6.0in,angle=0]{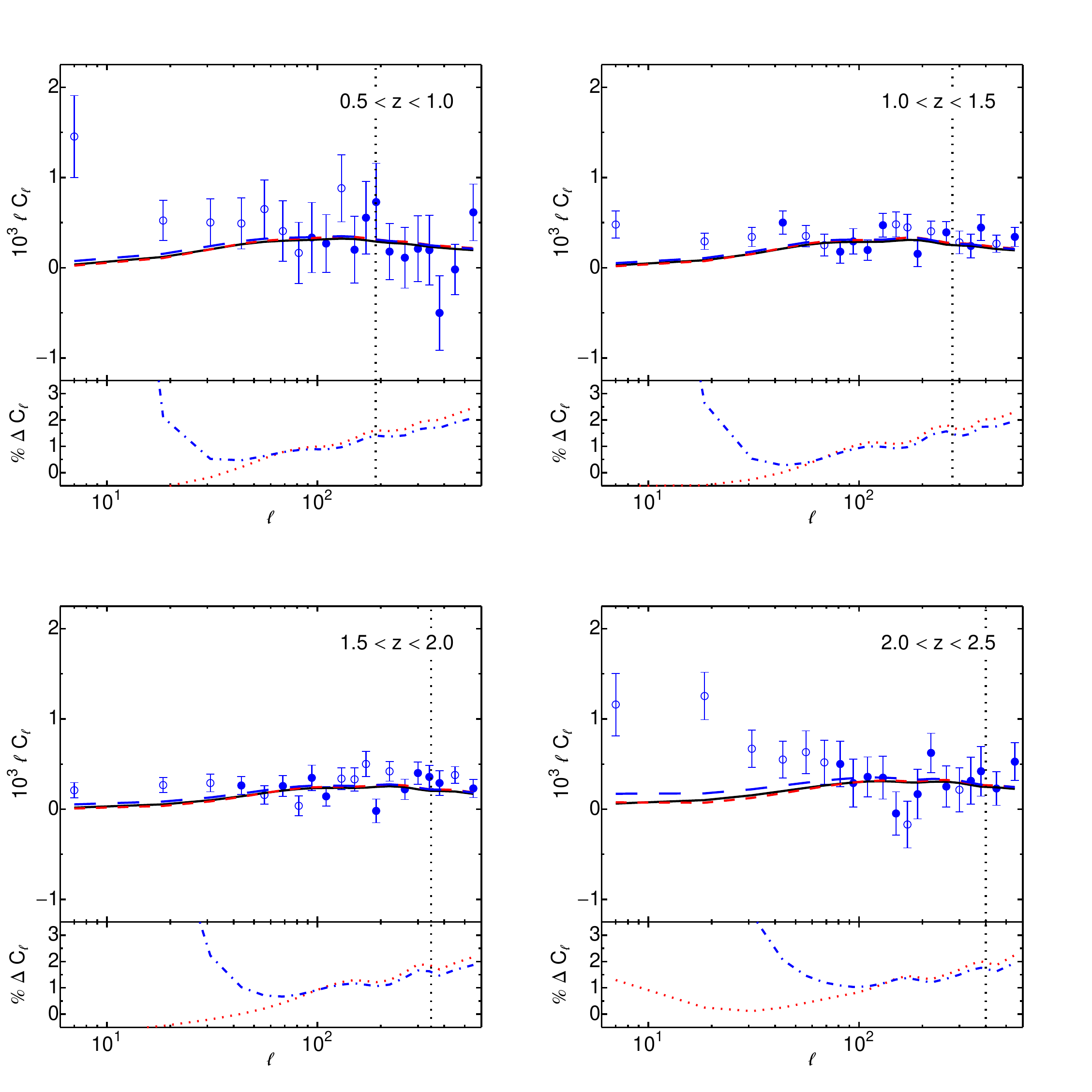}
	\caption{The angular power spectrum in the four redshift slices of quasars. The choice of symbols is the same as for LRGs in fig.\ \ref{fig1}. We note again that bins in each redshift slice that do not appear contaminated can still be dropped because their cross-power with another redshift slice is significantly contaminated and one cannot tell a priori which redshift slice is responsible for the contamination.}
\label{fig2}
\end{center}
\end{figure}

In figs. \ref{fig1} and \ref{fig2} we present the angular power spectra for LRGs and quasars. We mark the data points that are discarded in our analysis and the cutoff $\ell_{\rm max}$ that is imposed in each redshift slice. We also show the theoretical angular power spectrum for the best-fit value of $f_{\rm NL}$ (LRGs + quasars) with $68\%$ confidence limits. For all theoretical curves, we set the bias equal to their best-fit values and fix the background cosmology parameters corresponding to each value of $f_{\rm NL}$ separately.

To test the validity of the constant bias and non-linear fitting parameters assumption we perform the following test. We set $f_{\rm NL}$ to the best-fit and $\pm 1\sigma$ values for LRGs + quasars from table\ \ref{table3} and vary over all other free parameters using LRGs or quasars. In each case we find that values of the Gaussian bias shift by a few percent relative to those quoted in tables\ \ref{table1} and \ref{table2}. The $a$ parameters for LRGs shift within $1\sigma$, and are still consistent with zero at the $2\sigma$ level. This suggests that at least as a starting point, it is reasonable to set the bias and $a$ parameters to their best-fit values; with future data, however, it might be appropriate to vary over all free parameters. In addition, although there is evidence that non-Gaussianity also introduces a {\it scale-independent} correction to the bias (e.g., \cite{Giannantonio:2009ak}), this check shows that current data is not sensitive to the inclusion of that term.

\subsection{Constraints on ${\cal A}_{\rm NL}$ and $\alpha$}

General models of inflation with multiple fields or excited initial states can lead to deviations from the exact local ansatz. In addition to varying the scale-dependence $\alpha$ of the non-Gaussian halo bias, generic inflationary models also lead to a mass-dependent $f_{\rm NL}$, which we denote with ${\cal A}_{\rm NL}(b_{1}(M,z))$. An eventual goal would be to measure the amplitude ${\cal A}_{\rm NL}$ on different scales, or equivalently at different values of the bias (i.e. using a variety of LSS tracers with different masses or at many different redshifts). In either case, whether primordial non-Gaussianity (if non-zero) is scale-invariant or not, we should be able to learn a lot about the physics of inflation.

With the current LSS data, especially with large unknown systematics in the high-redshift data, it is difficult to constrain both the amplitude and the scale-dependence of the non-Gaussian correction to the halo bias. In particular, constraining $\alpha$ requires precise measurements of the power spectrum over a wide range of scales. Whereas on large scales we are dominated with various systematics in the data, small scales are highly non-linear. We therefore use different combinations of ${\cal A}_{\rm NL}$ and $\alpha$ with available data sets, to effectively constrain models of inflation.

We consider the following cases --- (i) constraints on ${\cal A}_{\rm NL}$ for $\alpha = 1.7$ using LRGs, quasars, and LRGs + quasars, (ii) constraints on ${\cal A}_{\rm NL}$ for $\alpha = 3$ using LRGs, quasars, and LRGs + quasars, and (iii) constraints on ${\cal A}_{\rm NL}$ and $\alpha$ using LRGs + quasars. In each case we assume that the halo mass for LRGs and quasars is uniform within a given redshift slice. We also note that it is not entirely consistent to combine data from LRGs and quasars for $\alpha \neq 2$ since ${\cal A}_{\rm NL}$ depends on the halo mass and redshift. However, since this dependence is expected to be weak \cite{Shandera:2010ei}, it is still useful to look at constraints from LRGs + quasars. The results of these studies are given in table\ \ref{table4}.

\begin{table}[!h]
\begin{center}
	\begin{tabular}{|c|l|l|}
		\hline
		Fixed parameter & \multicolumn{1}{|c|}{Data set} & \multicolumn{1}{|c|}{Constraints} \\
		($\alpha \ {\rm or} \ {\cal A}_{\rm NL}$) & & \\
		\hline
		& LRGs & ${\cal A}_{\rm NL} = -32^{+177}_{-180}$ \\
		$\alpha = 1.7$ & Quasars & ${\cal A}_{\rm NL} = 217^{+411}_{-410}$ \\
		& LRGs + quasars & ${\cal A}_{\rm NL} = -1^{+171}_{-171} $ \\
		\hline
		& LRGs & ${\cal A}_{\rm NL} = -3.5^{+10.8}_{-10.6}$ \\
		$\alpha = 3$ & Quasars & ${\cal A}_{\rm NL} = 1.8^{+12.3}_{-14.5}$ \\
		& LRGs + quasars & ${\cal A}_{\rm NL} = 0.9^{+4.1}_{-4.2}$ \\
		\hline
		--- & LRGs + quasars & See fig.\ \ref{fig3} \\
		\hline
	\end{tabular}
\caption{Constraints on ${\cal A}_{\rm NL}\left(k_{p} = 0.1 \ {\rm Mpc}^{-1}\right)$ and $\alpha$, with $68\%$ confidence limits, using different data sets.}
\label{table4}
\end{center}
\end{table}

From table\ \ref{table4} we first notice that at fixed $\alpha$ the constraints on ${\cal A}_{\rm NL}$ are more stringent for $\alpha = 3$ as compared to $\alpha = 1.7$. This can be understood as follows. For a given value of ${\cal A}_{\rm NL}\left(k_{p} = 0.1 \ {\rm Mpc}^{-1}\right)$, modifications to the power spectrum in the presence of primordial non-Gaussianity come in at the largest measured scales (i.e. at small $k$). This is no longer true when we allow for deviations from the local ansatz. In particular, as we increase the value of $\alpha$, non-Gaussian corrections become significant at smaller scales (close to matter-radiation equality) which are better measured, strongly constraining models of inflation that give $\alpha > 2$. On the other hand, for $0 < \alpha < 2$, non-Gaussian corrections are only significant at much larger scales, which are eventually limited by systematics. In figs.\ \ref{fig1} and \ref{fig2} we also show the percentage difference of the best-fit (LRGs + quasars) angular power spectrum for LRGs and quasars with respect to the Gaussian case (i.e. $f_{\rm NL} = 0$ or ${\cal A}_{\rm NL} = 0$) for $\alpha = 1.7$ and $\alpha = 3$.

\begin{figure}[!h]
\begin{center}
	\includegraphics[width=4.0in,angle=0]{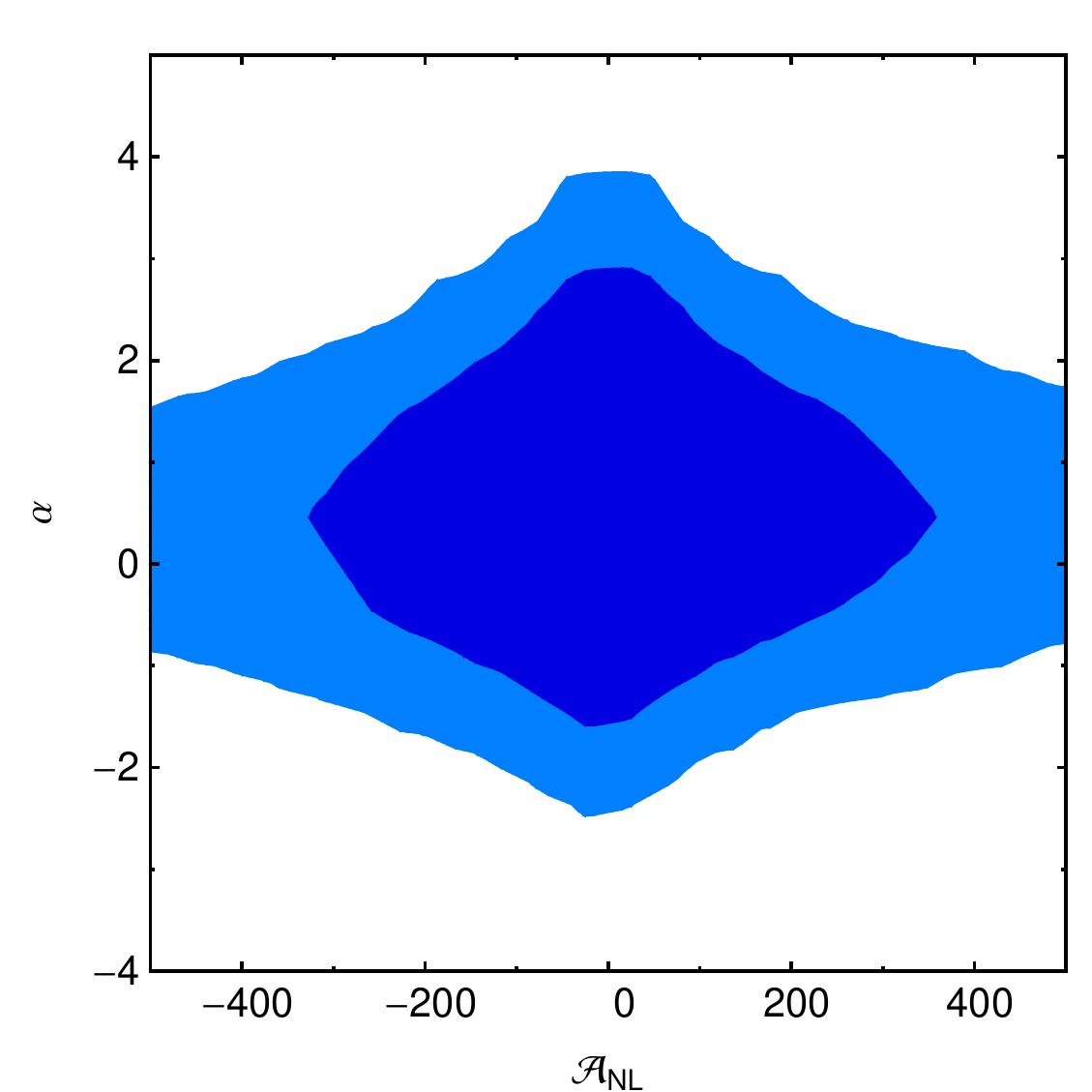}
	\caption{The $68\%$ (dark blue) and $95\%$ (light blue) confidence regions in the $(\alpha,{\cal A}_{\rm NL})$ parameter space using LRGs $+$ quasars.}
\label{fig3}
\end{center}
\end{figure}

Next we consider the case where we vary over both $\alpha$ and ${\cal A}_{\rm NL}$. Fig.\ \ref{fig3} shows the posterior probability distribution in the $(\alpha,{\cal A}_{\rm NL})$ parameter space using LRGs $+$ quasars. Note that there is an infinite degeneracy in $\alpha$ in the ${\cal A}_{\rm NL} = 0$ direction. The full marginalized upper limit on $\alpha$ is $2.0$ at the $95\%$ confidence level. This is in agreement with the CMB-galaxy cross-correlation constraints of \cite{Giannantonio:2013uqa} (also see \cite{Becker:2012je} for CMB constraints on the running of non-Gaussianity).

\section{Forecasts for surveys with minimal systematics}
\label{sec:fisher}

To compare the results of the previous section with what we might expect in the absence of any systematics, we show here Fisher matrix forecasts of constraints on $\alpha$ and ${\cal A}_{\rm NL}$ for the LRG and quasar data sets used earlier. We assume a WMAP9 + SN $\Lambda$CDM cosmology for the analysis in this section.

The Fisher information matrix evaluates how the likelihood, $\mathcal{L}$, of data for an observable depends on parameters $p_{\alpha}$ of a model for the data,
\bea
	F_{\alpha\beta} & \equiv & - \left\langle \frac{\partial^{2} \rm{ln} \mathcal{L}}{\partial p_{\alpha} \partial p_{\beta}} \right\rangle.
\eea
Here, the observables are the power spectra of LRGs and quasars. Assuming a negligible covariance between redshift bins,
\bea
	F_{\alpha\beta} & = & \frac{1}{(2\pi)^{2}} \sum_{l,m} \frac{\partial {\rm ln} \tilde{P}_{g}(k_{m},z_{l})}{\partial p_{\alpha}}\frac{\partial {\rm ln} \tilde{P}_{g}(k_{m},z_{l})}{\partial p_{\beta}} V_{l} \, k_{m}^{2} \, \Delta k,
\eea
where the sum is over redshift bins and wavenumber shells for each object, and
\bea
	V_{l} & = & \Omega_{\rm survey} \int_{z_{l}}^{z_{l} + \Delta z} \frac{{\rm d}V}{{\rm d}\Omega \, {\rm d}z} \, {\rm d}z
\eea
is the volume of each redshift bin within the survey. We assume that we have useable information for a range of scales between $k_{\rm min}$ corresponding to $\ell = 10$ at the mean redshift of each redshift slice and $k_{\rm max} = 0.1h \ {\rm Mpc}^{-1}$ and the survey volume is 11,000 square degrees. The power spectrum is evaluated at the median redshift in each bin and includes a shot noise term based on the actual number density of objects observed in each bin (after weighting by the probability that the object is of the desired type), $n_i$,
\bea
	\tilde{P}_{g} (k,z) & = & P_{g} (k,z_{\rm med}) + \frac{1}{n_{i}}.
\eea

\begin{figure}[!h]
\begin{center}
	\includegraphics[width=\textwidth,angle=0]{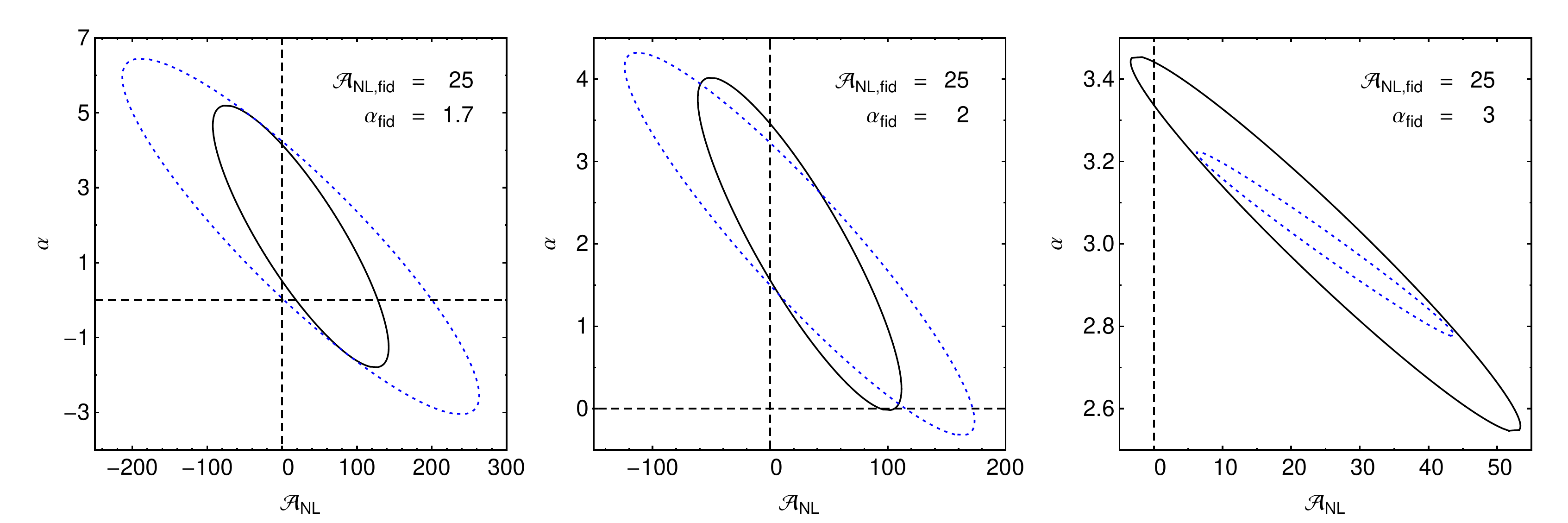}
	\caption{The one-sigma ellipses for LRGs (solid black) and quasars (dotted blue), with fiducial ${\cal A}_{\rm NL}(k_{p} = 0.1 \ {\rm Mpc}^{-1}) = 25$ and $\alpha = 1.7, \ 2, \ 3$. See also table\ \ref{table5}, which lists the results at the pivot points where ${\cal A}_{\rm NL}$ and $\alpha$ are uncorrelated for each choice of $\alpha_{\rm fid}$ and each data set (LRGs or quasars). }
\label{fig4}
\end{center}
\end{figure}

Observationally, the parameters ${\cal A}_{\rm NL}$ and $\alpha$ in the non-Gaussian bias capture the effects from a range of models. To give a sense of how Fisher forecasts compare to the actual utility of the data (once systematic errors have been considered), we use the parameterization given in eq.\ (\ref{eq:deltab}) with $k_{p} = 0.1 \ {\rm Mpc}^{-1}$. In addition, we calculate the best constrained wavenumber, $k_{\rm piv}$, where errors on ${\cal A}_{\rm NL}$ and $\alpha$ are uncorrelated (in terms of the original, fiducial scale $k_{\rm fid}$) \cite{Eisenstein:1998hr},
\bea
	k_{\rm piv,uncorr} & = & k_{\rm fid} \, {\rm Exp} \left( \frac{C_{\alpha{\cal A}_{\rm NL}}}{{\cal A}_{\rm NL}(k_{\rm fid}) \, C_{\alpha\alpha}} \right),
\eea
where $C_{ij}$ are the entries of the covariance matrix. In table\ \ref{table5} we show the Fisher matrix predictions for how well ideal versions of the data sets we have used would constrain non-Gaussianity. Fig.\ \ref{fig4} shows error ellipses to illustrate the degeneracy between ${\cal A}_{\rm NL}$ and $\alpha$ for different fiducial models using $k_{p} = 0.1 \ {\rm Mpc}^{-1}$, as adopted for the data analysis. Since we use the observed number densities for LRGs and quasars, rather than the expected number densities calculated from a mass function, the Fisher constraints from quasars can be weaker compared to those from LRGs. Further, the size of the errors tells us that quasars are shot noise limited, and hence the Fisher errors for quasars are closer to the results of \cite{Slosar:2008hx,Xia:2011hj,Xia:2010pe}.

\begin{table}[!h]
\begin{center}
	\begin{tabular}{|c|c|c|c||c|c|c|}
		\hline
		 \multirow{2}{*}{$\alpha_{\rm fid}$} &\multirow{2}{*}{Data set} & \multicolumn{2}{|c||}{At $k_{\rm piv} = 0.1 \ {\rm Mpc}^{-1}$} & \multicolumn{3}{|c|}{At $k_{\rm piv, uncorr}$} \\
		 \cline{3-7}
		 & & $\sigma({\cal A}_{\rm NL})$ & $\sigma(\alpha)$ & $k_{\rm piv, uncorr}$ (${\rm Mpc}^{-1}$) & $\sigma({\cal A}_{\rm NL})$ & $\sigma(\alpha)$ \\
\hline
		\multirow{2}{*}{1.7} & LRGs & 41.2 & 1.2 & 0.032 & 29.4 & 1.7 \\
		& Quasars & 69.9 & 1.4 & 0.017 & 40.7 & 1.9 \\
		\hline
		\multirow{2}{*}{2} & LRGs & 28.1 & 0.65 & 0.023 & 28.0 & 1.3 \\
		& Quasars & 37.0 & 0.58 & 0.005 & 37.6 & 1.7 \\
		\hline
		\multirow{2}{*}{3} & LRGs & 4.5 & 0.07 & 0.007 & 45.0 & 3.5 \\
		& Quasars & 1.8 & 0.02 & 0.004 & 56.4 & 4.8 \\
		\hline
	\end{tabular}
\caption{Fisher analysis results with ${\cal A}_{\rm NL, fid} = 25$ and $\alpha_{\rm fid} = 1.7, \ 2, \ 3$, for the LRG and quasar data sets, assuming systematic errors to be completely cleaned (and fixed cosmological parameters). For a pivot of $0.1 \ {\rm Mpc}^{-1}$, the marginalized one sigma values are shown. The last three columns show the pivot point in each data set where ${\cal A}_{\rm NL}$ and $\alpha$ are uncorrelated, and the one sigma uncertainties in that case. Notice that in this case, the level of non-Gaussianity at the scale $k = 0.1 \ {\rm Mpc}^{-1}$ is the same as in the first column for $\alpha_{\rm fid}=2$, smaller for $\alpha_{\rm fid}=3$, and larger for $\alpha_{\rm fid}=1.7$. This explains most of the trend for $\sigma(\mathcal{A}_{\rm NL})$ and $\sigma(\alpha)$ between the two choices of pivot.}
\label{table5}
\end{center}
\end{table}

\section{Analytic results}
\label{sec:theory}

The analysis in this paper is based on the motivation that different models of inflation can lead to forms of primordial non-Gaussianity that are more general than the exact local ansatz. We parameterize this in terms of a parameter $\alpha$, which can be different from two for a general model of non-Gaussianity. In this section we briefly review the analytic method to determine the non-Gaussian correction to the bias for a model of primordial non-Gaussianity different from the local ansatz, and use it to obtain a scaling of the amplitude ${\cal A}_{\rm NL}$ with the bias of the LSS tracer.

Using arguments based on the peak-background split formalism, it can be shown that the contribution from the bispectrum to the scale-dependent non-Gaussian bias is given by \cite{Matarrese:2008nc,Schmidt:2010,Desjacques:2011jb,Desjacques:2011mq,Baumann:2012bc}
\bea
	\Delta b_{\rm non-Gaussian} & = & 2 \frac{{\cal F}_{R}(k)}{{\cal M}(k,z)} \left\{ \left[ b_{1}(M,z) - p \right] \delta_{c} + \frac{{\rm d} \ln {\cal F}_{R}(k)}{{\rm d} \ln \sigma_{R}(z)} \right\}.
\label{eq:deltabtheory}
\eea
Here $R$ is the spatial smoothing scale related to the smoothing mass scale $M$ as $M = (4/3)\pi R^{3}\rho_{m}$, with $\rho_{m}$ being the matter energy density today. The function ${\cal M}(k,z)$ relates the matter density perturbations today to the primordial curvature perturbations, and is given by
\bea
	{\cal M}(k,z) & = & \frac{2}{5} \frac{k^{2}T(k)D(z)}{\Omega_{m}H_{0}^{2}}.
\eea
Since we compute statistics of the smoothed density field, it is also useful to define ${\cal M}_{R}(k,z) = {\cal M}(k,z) W_{R}(k)$, where $W_{R}(k) \equiv 3j_{1}(kR)/kR$ is the Fourier transform of a spherical top-hat window function of radius $R$. The smoothed variance of density fluctuations at redshift $z$ is then defined as
\bea
	\sigma^{2}_{R}(z) & = & \int \frac{{\rm d}^{3}k}{(2\pi)^{3}} \ {\cal M}_{R}^{2}(k,z) P_{\zeta}(k).
\eea
Finally, the redshift independent function ${\cal F}_{R}(k)$ is given by
\bea
	{\cal F}_{R}(k) & = & \frac{1}{4\sigma_{R}^{2}(z)P_{\zeta}(k)} \int \frac{{\rm d}^{3}q}{(2\pi)^{3}} \ {\cal M}_{R}(q,z) {\cal M}_{R}(\tilde{q},z) B_{\zeta}(q,\tilde{q},k),
\label{eq:defineF}
\eea
where $\tilde{q}^{2} = |\vec{k}+\vec{q}\,|^2$ and for the bispectrum $B_{\zeta}(k_{1},k_{2},k_{3})$ we can use any template, such as the bispectrum for a multi-field model of inflation or that for single-field inflation with a modified initial state. For the local ansatz, ${\cal F}_{R}(k)$ reduces approximately to $(3/5) f_{\rm NL}$ on large scales. In this limit, the scale-dependent correction to the bias is given by eq.\ (\ref{eq:deltabstd}).

To model the bispectra that would give a bias that goes as $1/k^{\alpha}$, we use two simple factorizable forms, one scale-independent (SI) and one scale-dependent (SD). These bispectra have two parameters, $f_{\rm NL}$ (or $f_{\rm NL}(k_p)$) and $n_{f}$, which will be related to the observational quantities ${\cal A}_{\rm NL}$ and $\alpha$. The bispectra we consider are
\bea
	B_{\zeta,{\rm SI}}(k_{1},k_{2},k_{3}) & = & \frac{6}{5} f_{\rm NL} \left[ \left( \frac{k_{1}}{k_{3}} \right)^{-n_{f}} \left( \frac{k_{2}}{k_{3}} \right)^{-n_{f}} P_{\zeta}(k_{1}) P_{\zeta}(k_{2}) + 2 \ {\rm perm.} \right], \\
	B_{\zeta,{\rm SD}}(k_{1},k_{2},k_{3}) & = & \frac{6}{5} f_{\rm NL}(k_{p}) \left[ \left( \frac{k_{1}}{k_{p}} \right)^{-n_{f}} \left( \frac{k_{2}}{k_{p}} \right)^{-n_{f}} P_{\zeta}(k_{1}) P_{\zeta}(k_{2}) + 2 \ {\rm perm.} \right].
\eea
The non-Gaussian bias will have a scale-dependence like $1/k^{2+n_{f}}$ in either case, so $\alpha = 2 + n_{f}$. However, the two bispectra will otherwise have different properties (e.g., they are constrained differently by the CMB).

For the above two forms of the bispectrum, the function ${\cal F}_{R}(k)$ in eq.\ (\ref{eq:defineF}) can be simplified as follows. For the scale-independent bispectrum,
\bea
	{\cal F}_{R,{\rm SI}}(k) & = & \frac{3f_{\rm NL}}{40\pi^{2}\sigma_{R}^{2}(z)} \int_{0}^{\infty} {\rm d}q \int_{-1}^{1} {\rm d}\mu \, q^{2} {\cal M}_{R}(q,z) {\cal M}_{R}(\tilde{q},z) P_{\zeta}(q) \nonumber \\
	& & \quad \quad \times \, \left[ \left( \frac{qk}{\tilde{q}^{2}} \right)^{-n_{f}} + \frac{P_{\zeta}(\tilde{q})}{P_{\zeta}(q)} \left( \frac{\tilde{q}k}{q^{2}} \right)^{-n_{f}} + \frac{P_{\zeta}(\tilde{q})}{P_{\zeta}(k)} \left( \frac{q\tilde{q}}{k^{2}} \right)^{-n_{f}} \right] \\
	& \xrightarrow[k \rightarrow 0]{} & \frac{3f_{\rm NL}}{40\pi^{2}\sigma_{R}^{2}(z)} \int_{0}^{\infty} {\rm d}q \int_{-1}^{1} {\rm d} \mu \, q^{2} {\cal M}_{R}^{2}(q,z) P_{\zeta}(q) \left[ 2\left( \frac{k}{q} \right)^{-n_{f}} \right],
\eea
where $\mu$ is the cosine of the angle between $\vec{k}$ and $\vec{q}$ and in the last line we have assumed that $n_{f} > - 1$. Similarly, for the scale-dependent case,
\bea
	{\cal F}_{R,{\rm SD}}(k) & = & \frac{3f_{\rm NL}}{40\pi^{2}\sigma_{R}^{2}(z)} \int_{0}^{\infty} {\rm d}q \int_{-1}^{1} {\rm d}\mu \, q^{2} {\cal M}_{R}(q,z) {\cal M}_{R}(\tilde{q},z) P_{\zeta}(q) \nonumber \\
	& & \quad \quad \times \, \left[ \left( \frac{qk}{k_{p}^{2}} \right)^{-n_{f}} + \frac{P_{\zeta}(\tilde{q})}{P_{\zeta}(q)} \left( \frac{\tilde{q}k}{k_{p}^{2}} \right)^{-n_{f}} + \frac{P_{\zeta}(\tilde{q})}{P_{\zeta}(k)} \left( \frac{q\tilde{q}}{k_{p}^{2}} \right)^{-n_{f}} \right] \\
	& \xrightarrow[k \rightarrow 0]{} & \frac{3f_{\rm NL}}{40\pi^{2}\sigma_{R}^{2}(z)} \int_{0}^{\infty} {\rm d}q \int_{-1}^{1} {\rm d} \mu \, q^{2} {\cal M}_{R}^{2}(q,z) P_{\zeta}(q) \left[ 2\left( \frac{qk}{k_{p}^{2}} \right)^{-n_{f}} \right].
\eea
The derivative term in eq.\ (\ref{eq:deltabtheory}) can be calculated using
\bea
	\frac{{\rm d} \ln {\cal F}_{R,{\rm SI}}(k)}{{\rm d} \ln \sigma_{R}(z)} & = & -2 + \frac{1}{{\cal F}_{R,{\rm SI}}(k)} \frac{3f_{\rm NL}}{40\pi^{2}\sigma_{R}(z)} \frac{{\rm d} R}{{\rm d} \sigma_{R}(z)} \int_{0}^{\infty} {\rm d}q \int_{-1}^{1} {\rm d}\mu \, q^{2} P_{\zeta}(q) \nonumber \\
	& & \quad \quad \times \, \left[ {\cal M}_{R}(q,z) \frac{{\rm d} {\cal M}_{R}(\tilde{q},z)}{{\rm d} R} + {\cal M}_{R}(\tilde{q},z) \frac{{\rm d} {\cal M}_{R}(q,z)}{{\rm d} R} \right] \nonumber \\
	& & \quad \quad \times \, \left[ \left( \frac{qk}{\tilde{q}^{2}} \right)^{-n_{f}} + \frac{P_{\zeta}(\tilde{q})}{P_{\zeta}(q)} \left( \frac{\tilde{q}k}{q^{2}} \right)^{-n_{f}} + \frac{P_{\zeta}(\tilde{q})}{P_{\zeta}(k)} \left( \frac{q\tilde{q}}{k^{2}} \right)^{-n_{f}} \right] \\
	& \xrightarrow[k \rightarrow 0]{} & -2 + \frac{1}{{\cal F}_{R,{\rm SI}}(k)} \frac{3f_{\rm NL}}{40\pi^{2}\sigma_{R}(z)} \frac{{\rm d} R}{{\rm d} \sigma_{R}(z)} \int_{0}^{\infty} {\rm d}q \int_{-1}^{1} {\rm d}\mu \, q^{2} P_{\zeta}(q) \nonumber \\
	& & \quad \quad \times \, \left[ 2{\cal M}_{R}(q,z) \frac{{\rm d} {\cal M}_{R}(q,z)}{{\rm d} R} \right] \left[ 2 \left( \frac{qk}{\tilde{q}^{2}} \right)^{-n_{f}} \right],
\eea
and similarly for the scale-dependent case.

\begin{figure}[!h]
\begin{center}
	\includegraphics[width=4.0in,angle=0]{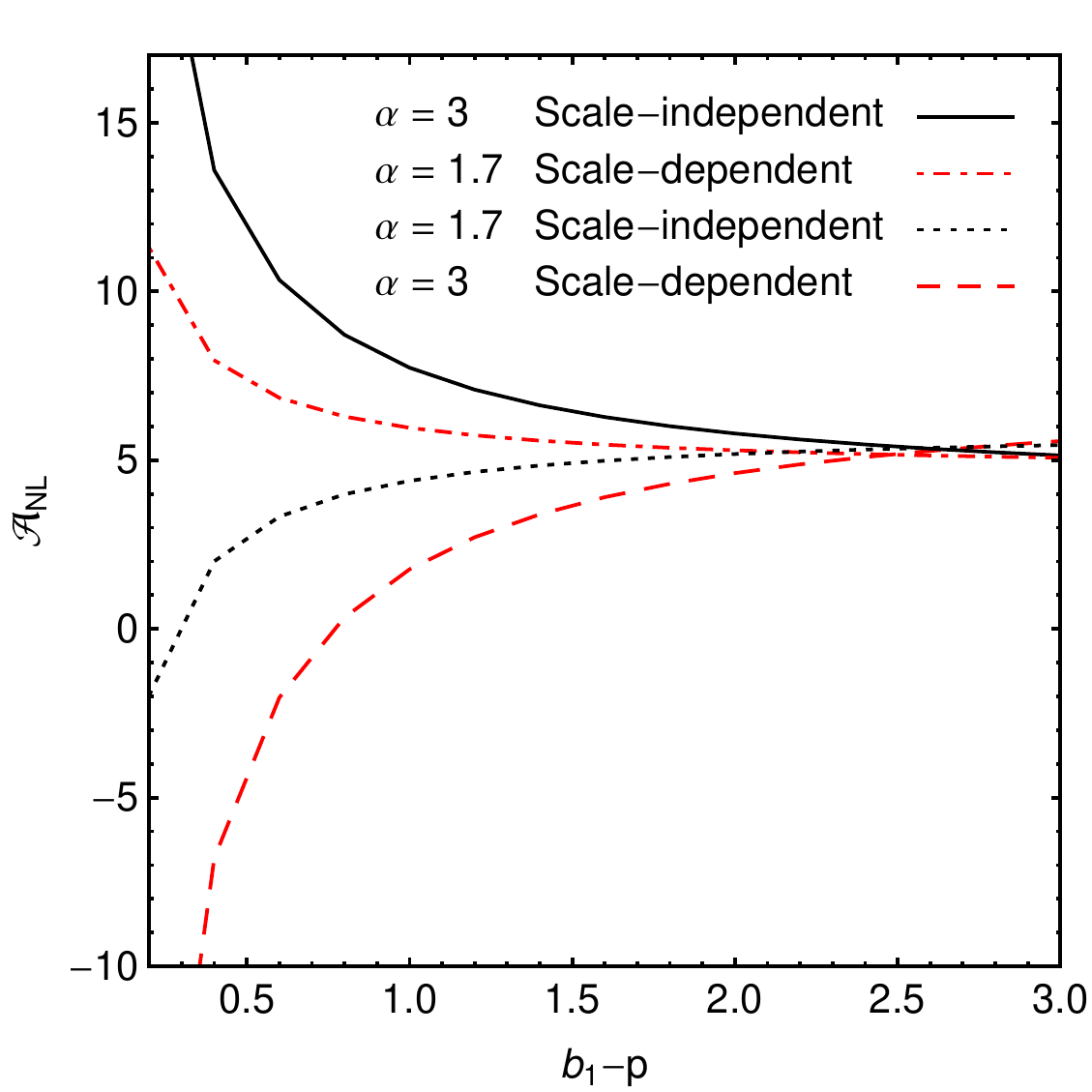}
	\caption{The amplitude of the non-Gaussian halo bias ${\cal A}_{\rm NL}\left( k_{p} = 0.1 \ {\rm Mpc}^{-1} \right)$ as a function of bias $(b_{1} - p)$ for $\alpha = 1.7$ and $\alpha = 3$, for the scale-independent and scale-dependent forms of the bispectrum. The curves are analytic, normalized to give ${\cal A}_{\rm NL} \approx 5$ at large $b_1-p$. For the local ansatz, and assuming spherical collapse, this choice corresponds to $f_{\rm NL} \approx 5/1.686$ in previous analyses.}
\label{fig5}
\end{center}
\end{figure}

We can now write the non-Gaussian bias in eq.\ (\ref{eq:deltabtheory}) in the form written earlier in eq.\ (\ref{eq:deltab}),
\bea
	\Delta b_{\rm non-Gaussian} & = & 3{\mathcal A}_{\rm NL}(b_{1}(M,z)) [b_{1}(M,z) - p] \frac{\Omega_{m}H_{0}^{2}}{k^{2}(k/k_{p})^{\alpha-2}T(k)D(z)},
\label{eq:deltabagain}
\eea
defining the amplitude ${\cal A}_{\rm NL}$ as
\bea
	{\cal A}_{\rm NL}(b_{1}(M,z)) & = & \frac{5}{3} {\cal F}_{R}(k) \left( \frac{k}{k_{p}} \right)^{\alpha - 2} \left\{ \delta_{c} + \frac{1}{[b_{1}(M,z) - p]} \frac{{\rm d} \ln {\cal F}_{R}(k)}{{\rm d} \ln \sigma_{R}(z)} \right\}.
\eea
This gives us the theoretically expected amplitude as a function of bias. We show ${\cal A}_{\rm NL}$ as a function of $b_{1} - p$ for $\alpha = 1.7$ and $\alpha = 3$, for the scale-independent and scale-dependent parameterizations of the bispectrum written earlier, in fig.\ \ref{fig5}. We take the pivot scale to be $k_{p} = 0.1 \ {\rm Mpc}^{-1}$ as before and normalize the curves to give ${\cal A}_{\rm NL} \approx 5$ at large $b_{1} - p$.

What this figure tells us is that it is possible to probe the scale-dependence of primordial non-Gaussianity using LSS. Using different tracers of LSS and with a good understanding of systematics at low $\ell$ we should be able to measure non-Gaussianity at different halo mass scales in the future.


\section{Discussion}
\label{sec:discussion}

LSS surveys contain a tremendous amount of cosmological information since they trace the full three-dimensional distribution of matter in the Universe. In particular, it has been noted that primordial non-Gaussianity affects the large scale clustering of virialized objects. General models of inflation, such as multi-field models and single-field inflation with a non-trivial initial state further lead to a generalized local ansatz for non-Gaussianity, modifying the scale dependence of the bias of dark matter halos away from the usual $f_{\rm NL}/k^{2}$ form.

In this paper, we use use a parameterization of the form ${\cal A}_{\rm NL}/k^{\alpha}$ to constrain both the amplitude and the scale-dependence of primordial non-Gaussianity. We find that photometric SDSS-III DR8 angular power spectrum data of LRGs and quasars is consistent with the local ansatz ($\alpha = 2$) at the $95\%$ confidence level on marginalizing over a range of ${\cal A}_{\rm NL}$ values. Constraints on the amplitude are weakened, compared to the local ansatz, for $\alpha < 2$ (e.g., we study $\alpha = 1.7$), while large values of $\alpha$ ($\alpha = 3$) are more tightly constrained, though still consistent with the current data.

Although we expect LSS to provide competitive or even better constraints on cosmological parameters compared to the CMB in the future, accessing this information is hard because of a lack of understanding of systematics in the data on large scales and the presence of non-linearities on small scales. In our data analysis we apply conservative cuts on systematics and on the non-linear scale. In order to assess how much we loose to contamination, we perform a Fisher matrix analysis using survey parameters of the same size as DR8, assuming perfectly clean data. We find that the errors on ${\cal A}_{\rm NL}$ and $\alpha$ reduce significantly, motivating a more thorough understanding of systematics in LSS data.

LSS surveys also open up the possibility of measuring the amplitude of primordial non-Gaussianity at different mass scales, or equivalently at different values of the bias. We present analytic results on the variation of ${\cal A}_{\rm NL}$ with bias $(b_{1} - p)$, for different forms of the bispectrum. This may be an important signature to look for in LSS data in the future. 


\acknowledgments It is a pleasure to thank Dragan Huterer for his Fisher code and for useful comments on this work. We also thank Ashley Disbrow for the quasar redshift distribution and Rachel Bean, Richard Holman, Anthony Pullen, Nic Ross, Uro\v{s} Seljak, An\v{z}e Slosar, and Michael Wood-Vasey for many helpful discussions. N. A. is supported by the McWilliams fellowship of the Bruce and Astrid McWilliams Center for Cosmology. The work of all three authors is also supported by the New Frontiers in Astronomy and Cosmology program at the John Templeton Foundation.

Funding for SDSS-III has been provided by the Alfred P. Sloan Foundation, the Participating Institutions, the National Science Foundation, and the U.S. Department of Energy Office of Science. The SDSS-III web site is http://www.sdss3.org/.

SDSS-III is managed by the Astrophysical Research Consortium for the Participating Institutions of the SDSS-III Collaboration including the University of Arizona, the Brazilian Participation Group, Brookhaven National Laboratory, University of Cambridge, Carnegie Mellon University, University of Florida, the French Participation Group, the German Participation Group, Harvard University, the Instituto de Astrofisica de Canarias, the Michigan State/Notre Dame/JINA Participation Group, Johns Hopkins University, Lawrence Berkeley National Laboratory, Max Planck Institute for Astrophysics, Max Planck Institute for Extraterrestrial Physics, New Mexico State University, New York University, Ohio State University, Pennsylvania State University, University of Portsmouth, Princeton University, the Spanish Participation Group, University of Tokyo, University of Utah, Vanderbilt University, University of Virginia, University of Washington, and Yale University.


\bibliography{references}

\providecommand{\href}[2]{#2}\begingroup\raggedright\begin{thebibliography}{100}

\bibitem{Ade:2013ydc}
P.~A.~R. Ade et~al., {\it {Planck 2013 Results. XXIV. Constraints on primordial
  non-Gaussianity}},  \href{http://xxx.lanl.gov/abs/1303.5084}{{\tt
  arXiv:1303.5084}}.

\bibitem{Hinshaw:2012aka}
G.~Hinshaw et~al., {\it {Nine-year Wilkinson Microwave Anisotropy Probe (WMAP)
  observations: Cosmological parameter results}},  {\em Astrophys.J.Suppl.}
  {\bf 208} (2013) 19, [\href{http://xxx.lanl.gov/abs/1212.5226}{{\tt
  arXiv:1212.5226}}].

\bibitem{Bennett:2012zja}
C.~L. Bennett et~al., {\it {Nine-year Wilkinson Microwave Anisotropy Probe
  (WMAP) observations: Final maps and results}},  {\em Astrophys.J.Suppl.} {\bf
  208} (2013) 20, [\href{http://xxx.lanl.gov/abs/1212.5225}{{\tt
  arXiv:1212.5225}}].

\bibitem{Dalal:2007cu}
N.~Dalal, O.~Dor\'{e}, D.~Huterer, and A.~Shirokov, {\it {The imprints of
  primordial non-Gaussianities on large-scale structure: Scale-dependent bias
  and abundance of virialized objects}},  {\em Phys.Rev.} {\bf D77} (2008)
  123514, [\href{http://xxx.lanl.gov/abs/0710.4560}{{\tt arXiv:0710.4560}}].

\bibitem{Grinstein:1986en}
B.~Grinstein and M.~B. Wise, {\it {Non-Gaussian fluctuations and the
  correlations of galaxies or rich clusters of galaxies}},  {\em Astrophys.J.}
  {\bf 310} (1986) 19--22.

\bibitem{Matarrese:1986et}
S.~Matarrese, F.~Lucchin, and S.~A. Bonometto, {\it {A path-integral approach
  to large-scale matter distribution originated by non-Gaussian fluctuations}},
   {\em Astrophys.J.} {\bf 310} (1986) L21--L26.

\bibitem{Ross:2012sx}
A.~J. Ross, W.~J. Percival, A.~Carnero, G.-b. Zhao, M.~Manera, et~al., {\it
  {The clustering of galaxies in SDSS-III DR9 Baryon Oscillation Spectroscopic
  Survey: Constraints on primordial non-Gaussianity}},
  \href{http://xxx.lanl.gov/abs/1208.1491}{{\tt arXiv:1208.1491}}.

\bibitem{Karagiannis:2013xea}
D.~Karagiannis, T.~Shanks, and N.~P. Ross, {\it {Search for primordial
  non-Gaussianity in the quasars of SDSS-III BOSS DR9}},
  \href{http://xxx.lanl.gov/abs/1310.6716}{{\tt arXiv:1310.6716}}.

\bibitem{Ho:2013lda}
S.~Ho, N.~Agarwal, A.~D. Myers, R.~Lyons, A.~Disbrow, et~al., {\it {Sloan
  Digital Sky Survey III photometric quasar clustering: Probing the initial
  conditions of the Universe using the largest volume}},
  \href{http://xxx.lanl.gov/abs/1311.2597}{{\tt arXiv:1311.2597}}.

\bibitem{York:2000gk}
D.~G. York et~al., {\it {The Sloan Digital Sky Survey: Technical summary}},
  {\em Astron.J.} {\bf 120} (2000) 1579--1587,
  [\href{http://xxx.lanl.gov/abs/astro-ph/0006396}{{\tt astro-ph/0006396}}].

\bibitem{Eisenstein:2011sa}
D.~J. Eisenstein et~al., {\it {SDSS-III: Massive spectroscopic surveys of the
  distant Universe, the milky way galaxy, and extra-solar planetary systems}},
  {\em Astron.J.} {\bf 142} (2011) 72,
  [\href{http://xxx.lanl.gov/abs/1101.1529}{{\tt arXiv:1101.1529}}].

\bibitem{Giannantonio:2013uqa}
T.~Giannantonio, A.~J. Ross, W.~J. Percival, R.~Crittenden, D.~Bacher, et~al.,
  {\it {Improved primordial non-Gaussianity constraints from measurements of
  galaxy clustering and the integrated Sachs-Wolfe effect}},  {\em Phys.Rev.}
  {\bf D89} (2014) 023511, [\href{http://xxx.lanl.gov/abs/1303.1349}{{\tt
  arXiv:1303.1349}}].

\bibitem{Cole:1989vx}
S.~Cole and N.~Kaiser, {\it {Biased clustering in the cold dark matter
  cosmogony}},  {\em Mon.Not.Roy.Astron.Soc.} {\bf 237} (1989) 1127--1146.

\bibitem{Gangui:1993tt}
A.~Gangui, F.~Lucchin, S.~Matarrese, and S.~Mollerach, {\it {The three-point
  correlation function of the cosmic microwave background in inflationary
  models}},  {\em Astrophys.J.} {\bf 430} (1994) 447--457,
  [\href{http://xxx.lanl.gov/abs/astro-ph/9312033}{{\tt astro-ph/9312033}}].

\bibitem{Verde:1999ij}
L.~Verde, L.-M. Wang, A.~Heavens, and M.~Kamionkowski, {\it {Large-scale
  structure, the cosmic microwave background, and primordial non-Gaussianity}},
   {\em Mon.Not.Roy.Astron.Soc.} {\bf 313} (2000) L141--L147,
  [\href{http://xxx.lanl.gov/abs/astro-ph/9906301}{{\tt astro-ph/9906301}}].

\bibitem{Komatsu:2001rj}
E.~Komatsu and D.~N. Spergel, {\it {Acoustic signatures in the primary
  microwave background bispectrum}},  {\em Phys.Rev.} {\bf D63} (2001) 063002,
  [\href{http://xxx.lanl.gov/abs/astro-ph/0005036}{{\tt astro-ph/0005036}}].

\bibitem{Matarrese:2008nc}
S.~Matarrese and L.~Verde, {\it {The effect of primordial non-Gaussianity on
  halo bias}},  {\em Astrophys.J.} {\bf 677} (2008) L77--L80,
  [\href{http://xxx.lanl.gov/abs/0801.4826}{{\tt arXiv:0801.4826}}].

\bibitem{Slosar:2008hx}
A.~Slosar, C.~Hirata, U.~Seljak, S.~Ho, and N.~Padmanabhan, {\it {Constraints
  on local primordial non-Gaussianity from large scale structure}},  {\em JCAP}
  {\bf 0808} (2008) 031, [\href{http://xxx.lanl.gov/abs/0805.3580}{{\tt
  arXiv:0805.3580}}].

\bibitem{Xia:2010yu}
J.-Q. Xia, M.~Viel, C.~Baccigalupi, G.~De~Zotti, S.~Matarrese, et~al., {\it
  {Primordial non-Gaussianity and the NRAO VLA Sky Survey}},  {\em
  Astrophys.J.} {\bf 717} (2010) L17,
  [\href{http://xxx.lanl.gov/abs/1003.3451}{{\tt arXiv:1003.3451}}].

\bibitem{Xia:2010pe}
J.-Q. Xia, A.~Bonaldi, C.~Baccigalupi, G.~De~Zotti, S.~Matarrese, et~al., {\it
  {Constraining primordial non-Gaussianity with high-redshift probes}},  {\em
  JCAP} {\bf 1008} (2010) 013, [\href{http://xxx.lanl.gov/abs/1007.1969}{{\tt
  arXiv:1007.1969}}].

\bibitem{Xia:2011hj}
J.-Q. Xia, C.~Baccigalupi, S.~Matarrese, L.~Verde, and M.~Viel, {\it
  {Constraints on primordial non-Gaussianity from large scale structure
  probes}},  {\em JCAP} {\bf 1108} (2011) 033,
  [\href{http://xxx.lanl.gov/abs/1104.5015}{{\tt arXiv:1104.5015}}].

\bibitem{Tashiro:2012wr}
H.~Tashiro and S.~Ho, {\it {Constraining primordial non-Gaussianity with
  CMB-21cm cross-correlations?}},
  \href{http://xxx.lanl.gov/abs/1205.0563}{{\tt arXiv:1205.0563}}.

\bibitem{Shandera:2010ei}
S.~Shandera, N.~Dalal, and D.~Huterer, {\it {A generalized local ansatz and its
  effect on halo bias}},  {\em JCAP} {\bf 1103} (2011) 017,
  [\href{http://xxx.lanl.gov/abs/1010.3722}{{\tt arXiv:1010.3722}}].

\bibitem{Desjacques:2011jb}
V.~Desjacques, D.~Jeong, and F.~Schmidt, {\it {Accurate predictions for the
  scale-dependent galaxy bias from primordial non-Gaussianity}},  {\em
  Phys.Rev.} {\bf D84} (2011) 061301,
  [\href{http://xxx.lanl.gov/abs/1105.3476}{{\tt arXiv:1105.3476}}].

\bibitem{Desjacques:2011mq}
V.~Desjacques, D.~Jeong, and F.~Schmidt, {\it {Non-Gaussian halo bias
  re-examined: Mass-dependent amplitude from the peak-background split and
  thresholding}},  {\em Phys.Rev.} {\bf D84} (2011) 063512,
  [\href{http://xxx.lanl.gov/abs/1105.3628}{{\tt arXiv:1105.3628}}].

\bibitem{Maldacena:2002vr}
J.~M. Maldacena, {\it {Non-Gaussian features of primordial fluctuations in
  single field inflationary models}},  {\em JHEP} {\bf 0305} (2003) 013,
  [\href{http://xxx.lanl.gov/abs/astro-ph/0210603}{{\tt astro-ph/0210603}}].

\bibitem{Creminelli:2004yq}
P.~Creminelli and M.~Zaldarriaga, {\it {Single field consistency relation for
  the 3-point function}},  {\em JCAP} {\bf 0410} (2004) 006,
  [\href{http://xxx.lanl.gov/abs/astro-ph/0407059}{{\tt astro-ph/0407059}}].

\bibitem{Creminelli:2011rh}
P.~Creminelli, G.~D'Amico, M.~Musso, and J.~Norena, {\it {The (not so) squeezed
  limit of the primordial 3-point function}},  {\em JCAP} {\bf 1111} (2011)
  038, [\href{http://xxx.lanl.gov/abs/1106.1462}{{\tt arXiv:1106.1462}}].

\bibitem{Pajer:2013ana}
E.~Pajer, F.~Schmidt, and M.~Zaldarriaga, {\it {The observed squeezed limit of
  cosmological three-point functions}},
  \href{http://xxx.lanl.gov/abs/1305.0824}{{\tt arXiv:1305.0824}}.

\bibitem{Linde:1996gt}
A.~D. Linde and V.~F. Mukhanov, {\it {Non-Gaussian isocurvature perturbations
  from inflation}},  {\em Phys.Rev.} {\bf D56} (1997) 535--539,
  [\href{http://xxx.lanl.gov/abs/astro-ph/9610219}{{\tt astro-ph/9610219}}].

\bibitem{Bernardeau:2002jy}
F.~Bernardeau and J.-P. Uzan, {\it {Non-Gaussianity in multifield inflation}},
  {\em Phys.Rev.} {\bf D66} (2002) 103506,
  [\href{http://xxx.lanl.gov/abs/hep-ph/0207295}{{\tt hep-ph/0207295}}].

\bibitem{Seery:2005gb}
D.~Seery and J.~E. Lidsey, {\it {Primordial non-Gaussianities from
  multiple-field inflation}},  {\em JCAP} {\bf 0509} (2005) 011,
  [\href{http://xxx.lanl.gov/abs/astro-ph/0506056}{{\tt astro-ph/0506056}}].

\bibitem{Rigopoulos:2005us}
G.~I. Rigopoulos, E.~P.~S. Shellard, and B.~J.~W. van Tent, {\it {Quantitative
  bispectra from multifield inflation}},  {\em Phys.Rev.} {\bf D76} (2007)
  083512, [\href{http://xxx.lanl.gov/abs/astro-ph/0511041}{{\tt
  astro-ph/0511041}}].

\bibitem{Vernizzi:2006ve}
F.~Vernizzi and D.~Wands, {\it {Non-gaussianities in two-field inflation}},
  {\em JCAP} {\bf 0605} (2006) 019,
  [\href{http://xxx.lanl.gov/abs/astro-ph/0603799}{{\tt astro-ph/0603799}}].

\bibitem{Battefeld:2006sz}
T.~Battefeld and R.~Easther, {\it {Non-Gaussianities in multi-field
  inflation}},  {\em JCAP} {\bf 0703} (2007) 020,
  [\href{http://xxx.lanl.gov/abs/astro-ph/0610296}{{\tt astro-ph/0610296}}].

\bibitem{Yokoyama:2007uu}
S.~Yokoyama, T.~Suyama, and T.~Tanaka, {\it {Primordial non-Gaussianity in
  multi-scalar slow-roll inflation}},  {\em JCAP} {\bf 0707} (2007) 013,
  [\href{http://xxx.lanl.gov/abs/0705.3178}{{\tt arXiv:0705.3178}}].

\bibitem{Yokoyama:2007dw}
S.~Yokoyama, T.~Suyama, and T.~Tanaka, {\it {Primordial non-Gaussianity in
  multi-scalar inflation}},  {\em Phys.Rev.} {\bf D77} (2008) 083511,
  [\href{http://xxx.lanl.gov/abs/0711.2920}{{\tt arXiv:0711.2920}}].

\bibitem{Sasaki:2008uc}
M.~Sasaki, {\it {Multi-brid inflation and non-Gaussianity}},  {\em
  Prog.Theor.Phys.} {\bf 120} (2008) 159--174,
  [\href{http://xxx.lanl.gov/abs/0805.0974}{{\tt arXiv:0805.0974}}].

\bibitem{Byrnes:2008wi}
C.~T. Byrnes, K.-Y. Choi, and L.~M.~H. Hall, {\it {Conditions for large
  non-Gaussianity in two-field slow-roll inflation}},  {\em JCAP} {\bf 0810}
  (2008) 008, [\href{http://xxx.lanl.gov/abs/0807.1101}{{\tt
  arXiv:0807.1101}}].

\bibitem{Byrnes:2010em}
C.~T. Byrnes and K.-Y. Choi, {\it {Review of local non-Gaussianity from
  multi-field inflation}},  {\em Adv.Astron.} {\bf 2010} (2010) 724525,
  [\href{http://xxx.lanl.gov/abs/1002.3110}{{\tt arXiv:1002.3110}}].

\bibitem{Kim:2010ud}
S.~A. Kim, A.~R. Liddle, and D.~Seery, {\it {Non-Gaussianity in axion Nflation
  models}},  {\em Phys.Rev.Lett.} {\bf 105} (2010) 181302,
  [\href{http://xxx.lanl.gov/abs/1005.4410}{{\tt arXiv:1005.4410}}].

\bibitem{Peterson:2010mv}
C.~M. Peterson and M.~Tegmark, {\it {Non-Gaussianity in two-field inflation}},
  {\em Phys.Rev.} {\bf D84} (2011) 023520,
  [\href{http://xxx.lanl.gov/abs/1011.6675}{{\tt arXiv:1011.6675}}].

\bibitem{Elliston:2011et}
J.~Elliston, D.~Mulryne, D.~Seery, and R.~Tavakol, {\it {Evolution of
  non-Gaussianity in multi-scalar field models}},  {\em Int.J.Mod.Phys.} {\bf
  A26} (2011) 3821--3832, [\href{http://xxx.lanl.gov/abs/1107.2270}{{\tt
  arXiv:1107.2270}}].

\bibitem{Dias:2012nf}
M.~Dias, J.~Frazer, and A.~R. Liddle, {\it {Multifield consequences for D-brane
  inflation}},  {\em JCAP} {\bf 1206} (2012) 020,
  [\href{http://xxx.lanl.gov/abs/1203.3792}{{\tt arXiv:1203.3792}}].

\bibitem{Lyth:2002my}
D.~H. Lyth, C.~Ungarelli, and D.~Wands, {\it {The Primordial density
  perturbation in the curvaton scenario}},  {\em Phys.Rev.} {\bf D67} (2003)
  023503, [\href{http://xxx.lanl.gov/abs/astro-ph/0208055}{{\tt
  astro-ph/0208055}}].

\bibitem{Ichikawa:2008iq}
K.~Ichikawa, T.~Suyama, T.~Takahashi, and M.~Yamaguchi, {\it {Non-Gaussianity,
  spectral index and tensor modes in mixed inflaton and curvaton models}},
  {\em Phys.Rev.} {\bf D78} (2008) 023513,
  [\href{http://xxx.lanl.gov/abs/0802.4138}{{\tt arXiv:0802.4138}}].

\bibitem{Beltran:2008aa}
M.~Beltran, {\it {Isocurvature, non-Gaussianity and the curvaton model}},  {\em
  Phys.Rev.} {\bf D78} (2008) 023530,
  [\href{http://xxx.lanl.gov/abs/0804.1097}{{\tt arXiv:0804.1097}}].

\bibitem{Chambers:2009ki}
A.~Chambers, S.~Nurmi, and A.~Rajantie, {\it {Non-Gaussianity from resonant
  curvaton decay}},  {\em JCAP} {\bf 1001} (2010) 012,
  [\href{http://xxx.lanl.gov/abs/0909.4535}{{\tt arXiv:0909.4535}}].

\bibitem{Byrnes:2009pe}
C.~T. Byrnes, S.~Nurmi, G.~Tasinato, and D.~Wands, {\it {Scale dependence of
  local $f_{\rm NL}$}},  {\em JCAP} {\bf 1002} (2010) 034,
  [\href{http://xxx.lanl.gov/abs/0911.2780}{{\tt arXiv:0911.2780}}].

\bibitem{Enqvist:2009ww}
K.~Enqvist, S.~Nurmi, O.~Taanila, and T.~Takahashi, {\it {Non-Gaussian
  fingerprints of self-interacting curvaton}},  {\em JCAP} {\bf 1004} (2010)
  009, [\href{http://xxx.lanl.gov/abs/0912.4657}{{\tt arXiv:0912.4657}}].

\bibitem{Alabidi:2010ba}
L.~Alabidi, K.~Malik, C.~T. Byrnes, and K.-Y. Choi, {\it {How the curvaton
  scenario, modulated reheating and an inhomogeneous end of inflation are
  related}},  {\em JCAP} {\bf 1011} (2010) 037,
  [\href{http://xxx.lanl.gov/abs/1002.1700}{{\tt arXiv:1002.1700}}].

\bibitem{Dvali:2003em}
G.~Dvali, A.~Gruzinov, and M.~Zaldarriaga, {\it {A new mechanism for generating
  density perturbations from inflation}},  {\em Phys.Rev.} {\bf D69} (2004)
  023505, [\href{http://xxx.lanl.gov/abs/astro-ph/0303591}{{\tt
  astro-ph/0303591}}].

\bibitem{Zaldarriaga:2003my}
M.~Zaldarriaga, {\it {Non-Gaussianities in models with a varying inflaton decay
  rate}},  {\em Phys.Rev.} {\bf D69} (2004) 043508,
  [\href{http://xxx.lanl.gov/abs/astro-ph/0306006}{{\tt astro-ph/0306006}}].

\bibitem{Suyama:2007bg}
T.~Suyama and M.~Yamaguchi, {\it {Non-Gaussianity in the modulated reheating
  scenario}},  {\em Phys.Rev.} {\bf D77} (2008) 023505,
  [\href{http://xxx.lanl.gov/abs/0709.2545}{{\tt arXiv:0709.2545}}].

\bibitem{Desjacques:2009jb}
V.~Desjacques and U.~Seljak, {\it {Signature of primordial non-Gaussianity of
  $\phi^3$-type in the mass function and bias of dark matter haloes}},  {\em
  Phys.Rev.} {\bf D81} (2010) 023006,
  [\href{http://xxx.lanl.gov/abs/0907.2257}{{\tt arXiv:0907.2257}}].

\bibitem{Smith:2011ub}
K.~M. Smith, S.~Ferraro, and M.~LoVerde, {\it {Halo clustering and
  $g_{NL}$-type primordial non-Gaussianity}},  {\em JCAP} {\bf 1203} (2012)
  032, [\href{http://xxx.lanl.gov/abs/1106.0503}{{\tt arXiv:1106.0503}}].

\bibitem{LoVerde:2011iz}
M.~LoVerde and K.~M. Smith, {\it {The non-Gaussian halo mass function with
  $f_{\rm NL}$, $g_{\rm NL}$ and $\tau_{\rm NL}$}},  {\em JCAP} {\bf 1108}
  (2011) 003, [\href{http://xxx.lanl.gov/abs/1102.1439}{{\tt
  arXiv:1102.1439}}].

\bibitem{Dias:2013rla}
M.~Dias, R.~H. Ribeiro, and D.~Seery, {\it {Scale-dependent bias from
  multiple-field inflation}},  {\em Phys.Rev.} {\bf D87} (2013) 107301,
  [\href{http://xxx.lanl.gov/abs/1303.6000}{{\tt arXiv:1303.6000}}].

\bibitem{Chen:2009we}
X.~Chen and Y.~Wang, {\it {Large non-Gaussianities with intermediate shapes
  from quasi-single field inflation}},  {\em Phys.Rev.} {\bf D81} (2010)
  063511, [\href{http://xxx.lanl.gov/abs/0909.0496}{{\tt arXiv:0909.0496}}].

\bibitem{Chen:2009zp}
X.~Chen and Y.~Wang, {\it {Quasi-single field inflation and
  non-Gaussianities}},  {\em JCAP} {\bf 1004} (2010) 027,
  [\href{http://xxx.lanl.gov/abs/0911.3380}{{\tt arXiv:0911.3380}}].

\bibitem{Chen:2006nt}
X.~Chen, M.-x. Huang, S.~Kachru, and G.~Shiu, {\it {Observational signatures
  and non-Gaussianities of general single field inflation}},  {\em JCAP} {\bf
  0701} (2007) 002, [\href{http://xxx.lanl.gov/abs/hep-th/0605045}{{\tt
  hep-th/0605045}}].

\bibitem{Holman:2007na}
R.~Holman and A.~J. Tolley, {\it {Enhanced non-Gaussianity from excited initial
  states}},  {\em JCAP} {\bf 0805} (2008) 001,
  [\href{http://xxx.lanl.gov/abs/0710.1302}{{\tt arXiv:0710.1302}}].

\bibitem{Chen:2008wn}
X.~Chen, R.~Easther, and E.~A. Lim, {\it {Generation and characterization of
  large non-Gaussianities in single field inflation}},  {\em JCAP} {\bf 0804}
  (2008) 010, [\href{http://xxx.lanl.gov/abs/0801.3295}{{\tt
  arXiv:0801.3295}}].

\bibitem{Meerburg:2009ys}
P.~D. Meerburg, J.~P. van~der Schaar, and P.~S. Corasaniti, {\it {Signatures of
  initial state modifications on bispectrum statistics}},  {\em JCAP} {\bf
  0905} (2009) 018, [\href{http://xxx.lanl.gov/abs/0901.4044}{{\tt
  arXiv:0901.4044}}].

\bibitem{Agullo:2010ws}
I.~Agullo and L.~Parker, {\it {Non-Gaussianities and the stimulated creation of
  quanta in the inflationary universe}},  {\em Phys.Rev.} {\bf D83} (2011)
  063526, [\href{http://xxx.lanl.gov/abs/1010.5766}{{\tt arXiv:1010.5766}}].

\bibitem{Ashoorioon:2010xg}
A.~Ashoorioon and G.~Shiu, {\it {A note on calm excited states of inflation}},
  {\em JCAP} {\bf 1103} (2011) 025,
  [\href{http://xxx.lanl.gov/abs/1012.3392}{{\tt arXiv:1012.3392}}].

\bibitem{Ganc:2011dy}
J.~Ganc, {\it {Calculating the local-type $f_{\rm NL}$ for slow-roll inflation
  with a non-vacuum initial state}},  {\em Phys.Rev.} {\bf D84} (2011) 063514,
  [\href{http://xxx.lanl.gov/abs/1104.0244}{{\tt arXiv:1104.0244}}].

\bibitem{Kundu:2011sg}
S.~Kundu, {\it {Inflation with general initial conditions for scalar
  perturbations}},  {\em JCAP} {\bf 1202} (2012) 005,
  [\href{http://xxx.lanl.gov/abs/1110.4688}{{\tt arXiv:1110.4688}}].

\bibitem{Chialva:2011hc}
D.~Chialva, {\it {Signatures of very high energy physics in the squeezed limit
  of the bispectrum}},  {\em JCAP} {\bf 1210} (2012) 037,
  [\href{http://xxx.lanl.gov/abs/1108.4203}{{\tt arXiv:1108.4203}}].

\bibitem{Agarwal:2012mq}
N.~Agarwal, R.~Holman, A.~J. Tolley, and J.~Lin, {\it {Effective field theory
  and non-Gaussianity from general inflationary states}},  {\em JHEP} {\bf
  1305} (2013) 085, [\href{http://xxx.lanl.gov/abs/1212.1172}{{\tt
  arXiv:1212.1172}}].

\bibitem{Flauger:2013hra}
R.~Flauger, D.~Green, and R.~A. Porto, {\it {On squeezed limits in single-field
  inflation. Part I}},  {\em JCAP} {\bf 1308} (2013) 032,
  [\href{http://xxx.lanl.gov/abs/1303.1430}{{\tt arXiv:1303.1430}}].

\bibitem{Aravind:2013lra}
A.~Aravind, D.~Lorshbough, and S.~Paban, {\it {Non-Gaussianity from excited
  initial inflationary states}},  {\em JHEP} {\bf 1307} (2013) 076,
  [\href{http://xxx.lanl.gov/abs/1303.1440}{{\tt arXiv:1303.1440}}].

\bibitem{Ashoorioon:2013eia}
A.~Ashoorioon, K.~Dimopoulos, M.~M. Sheikh-Jabbari, and G.~Shiu, {\it
  {Reconciliation of high energy scale models of inflation with Planck}},
  \href{http://xxx.lanl.gov/abs/1306.4914}{{\tt arXiv:1306.4914}}.

\bibitem{Kundu:2013gha}
S.~Kundu, {\it {Non-Gaussianity consistency relations, initial states and
  back-reaction}},  \href{http://xxx.lanl.gov/abs/1311.1575}{{\tt
  arXiv:1311.1575}}.

\bibitem{Ganc:2012ae}
J.~Ganc and E.~Komatsu, {\it {Scale-dependent bias of galaxies and $\mu$-type
  distortion of the cosmic microwave background spectrum from single-field
  inflation with a modified initial state}},  {\em Phys.Rev.} {\bf D86} (2012)
  023518, [\href{http://xxx.lanl.gov/abs/1204.4241}{{\tt arXiv:1204.4241}}].

\bibitem{Agullo:2012cs}
I.~Agullo and S.~Shandera, {\it {Large non-Gaussian halo bias from single field
  inflation}},  {\em JCAP} {\bf 1209} (2012) 007,
  [\href{http://xxx.lanl.gov/abs/1204.4409}{{\tt arXiv:1204.4409}}].

\bibitem{Becker:2010hx}
A.~Becker, D.~Huterer, and K.~Kadota, {\it {Scale-dependent non-Gaussianity as
  a generalization of the local model}},  {\em JCAP} {\bf 1101} (2011) 006,
  [\href{http://xxx.lanl.gov/abs/1009.4189}{{\tt arXiv:1009.4189}}].

\bibitem{Sefusatti:2012ye}
E.~Sefusatti, J.~R. Fergusson, X.~Chen, and E.~P.~S. Shellard, {\it {Effects
  and detectability of quasi-single field inflation in the large-scale
  structure and cosmic microwave background}},  {\em JCAP} {\bf 1208} (2012)
  033, [\href{http://xxx.lanl.gov/abs/1204.6318}{{\tt arXiv:1204.6318}}].

\bibitem{Norena:2012yi}
J.~Norena, L.~Verde, G.~Barenboim, and C.~Bosch, {\it {Prospects for
  constraining the shape of non-Gaussianity with the scale-dependent bias}},
  {\em JCAP} {\bf 1208} (2012) 019,
  [\href{http://xxx.lanl.gov/abs/1204.6324}{{\tt arXiv:1204.6324}}].

\bibitem{Becker:2012yr}
A.~Becker, D.~Huterer, and K.~Kadota, {\it {Constraining scale-dependent
  non-Gaussianity with future large-scale structure and the CMB}},  {\em JCAP}
  {\bf 1212} (2012) 034, [\href{http://xxx.lanl.gov/abs/1206.6165}{{\tt
  arXiv:1206.6165}}].

\bibitem{Becker:2012je}
A.~Becker and D.~Huterer, {\it {First constraints on the running of
  non-Gaussianity}},  {\em Phys.Rev.Lett.} {\bf 109} (2012) 121302,
  [\href{http://xxx.lanl.gov/abs/1207.5788}{{\tt arXiv:1207.5788}}].

\bibitem{Nelson:2012sb}
E.~Nelson and S.~Shandera, {\it {Statistical naturalness and non-Gaussianity in
  a finite Universe}},  {\em Phys.Rev.Lett.} {\bf 110} (2013) 131301,
  [\href{http://xxx.lanl.gov/abs/1212.4550}{{\tt arXiv:1212.4550}}].

\bibitem{Nurmi:2013xv}
S.~Nurmi, C.~T. Byrnes, and G.~Tasinato, {\it {A non-Gaussian landscape}},
  {\em JCAP} {\bf 1306} (2013) 004,
  [\href{http://xxx.lanl.gov/abs/1301.3128}{{\tt arXiv:1301.3128}}].

\bibitem{LoVerde:2013xka}
M.~LoVerde, E.~Nelson, and S.~Shandera, {\it {Non-Gaussian mode coupling and
  the statistical cosmological principle}},  {\em JCAP} {\bf 1306} (2013) 024,
  [\href{http://xxx.lanl.gov/abs/1303.3549}{{\tt arXiv:1303.3549}}].

\bibitem{Byrnes:2013qjy}
C.~T. Byrnes, S.~Nurmi, G.~Tasinato, and D.~Wands, {\it {Implications of the
  Planck bispectrum constraints for the primordial trispectrum}},  {\em
  Europhys.Lett.} {\bf 103} (2013) 19001,
  [\href{http://xxx.lanl.gov/abs/1306.2370}{{\tt arXiv:1306.2370}}].

\bibitem{Bramante:2013moa}
J.~Bramante, J.~Kumar, E.~Nelson, and S.~Shandera, {\it {Cosmic variance of the
  spectral index from mode coupling}},  {\em JCAP} {\bf 1311} (2013) 021,
  [\href{http://xxx.lanl.gov/abs/1307.5083}{{\tt arXiv:1307.5083}}].

\bibitem{Aihara:2011sj}
H.~Aihara et~al., {\it {The eighth data release of the Sloan Digital Sky
  Survey: First data from SDSS-III}},  {\em Astrophys.J.Suppl.} {\bf 193}
  (2011) 29, [\href{http://xxx.lanl.gov/abs/1101.1559}{{\tt arXiv:1101.1559}}].

\bibitem{Ross:2011cz}
A.~J. Ross, S.~Ho, A.~J. Cuesta, R.~Tojeiro, W.~J. Percival, et~al., {\it
  {Ameliorating systematic uncertainties in the angular clustering of galaxies:
  A study using SDSS-III}},  {\em Mon.Not.Roy.Astron.Soc.} {\bf 417} (2011)
  1350, [\href{http://xxx.lanl.gov/abs/1105.2320}{{\tt arXiv:1105.2320}}].

\bibitem{Ho:2012vy}
S.~Ho, A.~Cuesta, H.-J. Seo, R.~de~Putter, A.~J. Ross, et~al., {\it {Clustering
  of Sloan Digital Sky Survey III photometric luminous galaxies: The
  measurement, systematics and cosmological implications}},  {\em Astrophys.J.}
  {\bf 761} (2012) 14, [\href{http://xxx.lanl.gov/abs/1201.2137}{{\tt
  arXiv:1201.2137}}].

\bibitem{Lewis:2002ah}
A.~Lewis and S.~Bridle, {\it {Cosmological parameters from CMB and other data:
  A Monte-Carlo approach}},  {\em Phys.Rev.} {\bf D66} (2002) 103511,
  [\href{http://xxx.lanl.gov/abs/astro-ph/0205436}{{\tt astro-ph/0205436}}].

\bibitem{Lewis:1999bs}
A.~Lewis, A.~Challinor, and A.~Lasenby, {\it Efficient computation of {CMB}
  anisotropies in closed {FRW} models},  {\em Astrophys. J.} {\bf 538} (2000)
  473--476, [\href{http://xxx.lanl.gov/abs/astro-ph/9911177}{{\tt
  astro-ph/9911177}}].

\bibitem{Smith:2002dz}
R.~E. Smith et~al., {\it {Stable clustering, the halo model and nonlinear
  cosmological power spectra}},  {\em Mon.Not.Roy.Astron.Soc.} {\bf 341} (2003)
  1311, [\href{http://xxx.lanl.gov/abs/astro-ph/0207664}{{\tt
  astro-ph/0207664}}].

\bibitem{ST1}
R.~K. Sheth and G.~Tormen, {\it {Large-scale bias and the peak background
  split}},  {\em Mon.Not.Roy.Astron.Soc.} {\bf 308} (1999) 119--126,
  [\href{http://xxx.lanl.gov/abs/astro-ph/9901122}{{\tt astro-ph/9901122}}].

\bibitem{ST2}
R.~K. Sheth, H.~J. Mo, and G.~Tormen, {\it {Ellipsoidal collapse and an
  improved model for the number and spatial distribution of dark matter
  haloes}},  {\em Mon.Not.Roy.Astron.Soc.} {\bf 323} (2001) 1--12,
  [\href{http://xxx.lanl.gov/abs/astro-ph/9907024}{{\tt astro-ph/9907024}}].

\bibitem{Padmanabhan:2006ku}
N.~Padmanabhan et~al., {\it {The clustering of luminous red galaxies in the
  Sloan Digital Sky Survey imaging data}},  {\em Mon.Not.Roy.Astron.Soc.} {\bf
  378} (2007) 852--872, [\href{http://xxx.lanl.gov/abs/astro-ph/0605302}{{\tt
  astro-ph/0605302}}].

\bibitem{dePutter:2012sh}
R.~de~Putter, O.~Mena, E.~Giusarma, S.~Ho, A.~Cuesta, et~al., {\it {New
  neutrino mass bounds from Sloan Digital Sky Survey III Data Release 8
  photometric luminous galaxies}},  {\em Astrophys.J.} {\bf 761} (2012) 12,
  [\href{http://xxx.lanl.gov/abs/1201.1909}{{\tt arXiv:1201.1909}}].

\bibitem{Limber:1954zz}
D.~N. Limber, {\it {The analysis of counts of the extragalactic nebulae in
  terms of a fluctuating density field. II}},  {\em Astrophys.J.} {\bf 119}
  (1954) 655.

\bibitem{Seljak:1997ep}
U.~Seljak, {\it {Weak lensing reconstruction and power spectrum estimation:
  Minimum variance methods}},  {\em Astrophys.J.} {\bf 506} (1997) 64,
  [\href{http://xxx.lanl.gov/abs/astro-ph/9711124}{{\tt astro-ph/9711124}}].

\bibitem{Tegmark:1997yq}
M.~Tegmark, A.~J. Hamilton, M.~A. Strauss, M.~S. Vogeley, and A.~S. Szalay,
  {\it {Measuring the galaxy power spectrum with future redshift surveys}},
  {\em Astrophys.J.} {\bf 499} (1998) 555--576,
  [\href{http://xxx.lanl.gov/abs/astro-ph/9708020}{{\tt astro-ph/9708020}}].

\bibitem{Padmanabhan:2002yv}
N.~Padmanabhan, U.~Seljak, and U.~L. Pen, {\it {Mining weak lensing surveys}},
  {\em New Astron.} {\bf 8} (2003) 581--603,
  [\href{http://xxx.lanl.gov/abs/astro-ph/0210478}{{\tt astro-ph/0210478}}].

\bibitem{Amanullah:2010vv}
R.~Amanullah, C.~Lidman, D.~Rubin, G.~Aldering, P.~Astier, et~al., {\it
  {Spectra and light curves of six type Ia supernovae at $0.511 < z < 1.12$ and
  the Union2 compilation}},  {\em Astrophys.J.} {\bf 716} (2010) 712--738,
  [\href{http://xxx.lanl.gov/abs/1004.1711}{{\tt arXiv:1004.1711}}].

\bibitem{Gunn:2006tw}
J.~E. Gunn et~al., {\it {The 2.5 m telescope of the Sloan Digital Sky Survey}},
   {\em Astron.J.} {\bf 131} (2006) 2332--2359,
  [\href{http://xxx.lanl.gov/abs/astro-ph/0602326}{{\tt astro-ph/0602326}}].

\bibitem{Gunn:1998vh}
J.~E. Gunn et~al., {\it {The Sloan Digital Sky Survey photometric camera}},
  {\em Astron.J.} {\bf 116} (1998) 3040,
  [\href{http://xxx.lanl.gov/abs/astro-ph/9809085}{{\tt astro-ph/9809085}}].

\bibitem{Fukugita:1996qt}
M.~Fukugita, T.~Ichikawa, J.~E. Gunn, M.~Doi, K.~Shimasaku, et~al., {\it {The
  Sloan Digital Sky Survey photometric system}},  {\em Astron.J.} {\bf 111}
  (1996) 1748.

\bibitem{Smith:2002pca}
J.~A. Smith et~al., {\it {The ugriz standard star system}},  {\em Astron.J.}
  {\bf 123} (2002) 2121--2144,
  [\href{http://xxx.lanl.gov/abs/astro-ph/0201143}{{\tt astro-ph/0201143}}].

\bibitem{Pier:2002iq}
J.~R. Pier, J.~A. Munn, R.~B. Hindsley, G.~S. Hennessy, S.~M. Kent, et~al.,
  {\it {Astrometric calibration of the Sloan Digital Sky Survey}},  {\em
  Astron.J.} {\bf 125} (2003) 1559,
  [\href{http://xxx.lanl.gov/abs/astro-ph/0211375}{{\tt astro-ph/0211375}}].

\bibitem{Lupton:2001zb}
R.~Lupton et~al., {\it {The SDSS imaging pipelines}},  {\em ASP Conf.Ser.} {\bf
  238} (2001) 269--278, [\href{http://xxx.lanl.gov/abs/astro-ph/0101420}{{\tt
  astro-ph/0101420}}].

\bibitem{Padmanabhan:2007zd}
N.~Padmanabhan, D.~J. Schlegel, D.~P. Finkbeiner, J.~C. Barentine, M.~R.
  Blanton, et~al., {\it {An improved photometric calibration of the Sloan
  Digital Sky Survey imaging data}},  {\em Astrophys.J.} {\bf 674} (2008)
  1217--1233, [\href{http://xxx.lanl.gov/abs/astro-ph/0703454}{{\tt
  astro-ph/0703454}}].

\bibitem{Dawson:2012va}
K.~S. Dawson et~al., {\it {The Baryon Oscillation Spectroscopic Survey of
  SDSS-III}},  {\em Astron.J.} {\bf 145} (2013) 10,
  [\href{http://xxx.lanl.gov/abs/1208.0022}{{\tt arXiv:1208.0022}}].

\bibitem{Bolton:2012hz}
A.~S. Bolton et~al., {\it {Spectral classification and redshift measurement for
  the SDSS-III Baryon Oscillation Spectroscopic Survey}},  {\em Astron.J.} {\bf
  144} (2012) 144, [\href{http://xxx.lanl.gov/abs/1207.7326}{{\tt
  arXiv:1207.7326}}].

\bibitem{Smee:2012wd}
S.~Smee, J.~E. Gunn, A.~Uomoto, N.~Roe, D.~Schlegel, et~al., {\it {The
  multi-object, fiber-fed spectrographs for SDSS and the Baryon Oscillation
  Spectroscopic Survey}},  {\em Astron.J.} {\bf 126} (2013) 32,
  [\href{http://xxx.lanl.gov/abs/1208.2233}{{\tt arXiv:1208.2233}}].

\bibitem{White:2010ed}
M.~White, M.~Blanton, A.~Bolton, D.~Schlegel, J.~Tinker, et~al., {\it {The
  clustering of massive galaxies at $z \sim 0.5$ from the first semester of
  BOSS data}},  {\em Astrophys.J.} {\bf 728} (2011) 126,
  [\href{http://xxx.lanl.gov/abs/1010.4915}{{\tt arXiv:1010.4915}}].

\bibitem{Agarwal:2013ajb}
N.~Agarwal, S.~Ho, A.~D. Myers, H.-J. Seo, A.~J. Ross, et~al., {\it
  {Characterizing unknown systematics in large scale structure surveys}},
  \href{http://xxx.lanl.gov/abs/1309.2954}{{\tt arXiv:1309.2954}}.

\bibitem{Huterer:2012zs}
D.~Huterer, C.~E. Cunha, and W.~Fang, {\it {Calibration errors unleashed:
  Effects on cosmological parameters and requirements for large-scale structure
  surveys}},  {\em Mon.Not.Roy.Astron.Soc.} {\bf 432} (2013) 2945--2961,
  [\href{http://xxx.lanl.gov/abs/1211.1015}{{\tt arXiv:1211.1015}}].

\bibitem{Pullen:2012rd}
A.~R. Pullen and C.~M. Hirata, {\it {Systematic effects in large-scale angular
  power spectra of photometric quasars and implications for constraining
  primordial non-Gaussianity}},  {\em Publications of the Astronomical Society
  of the Pacific} {\bf 125} (2013) 705--718,
  [\href{http://xxx.lanl.gov/abs/1212.4500}{{\tt arXiv:1212.4500}}].

\bibitem{Hernandez-Monteagudo:2013vwa}
C.~Hern\'{a}ndez-Monteagudo, A.~Ross, A.~Cuesta, R.~G\'{e}nova-Santos,
  F.~Prada, et~al., {\it {The SDSS-III Baryonic Oscillation Spectroscopic
  Survey: Constraints on the integrated Sachs-Wolfe effect}},
  \href{http://xxx.lanl.gov/abs/1303.4302}{{\tt arXiv:1303.4302}}.

\bibitem{Leistedt:2013gfa}
B.~Leistedt, H.~V. Peiris, D.~J. Mortlock, A.~Benoit-L\'{e}vy, and A.~Pontzen,
  {\it {Estimating the large-scale angular power spectrum in the presence of
  systematics: A case study of Sloan Digital Sky Survey quasars}},  {\em
  Mon.Not.Roy.Astron.Soc.} {\bf 435} (2013) 1857,
  [\href{http://xxx.lanl.gov/abs/1306.0005}{{\tt arXiv:1306.0005}}].

\bibitem{Reid:2010vc}
B.~A. Reid, L.~Verde, K.~Dolag, S.~Matarrese, and L.~Moscardini, {\it
  {Non-Gaussian halo assembly bias}},  {\em JCAP} {\bf 1007} (2010) 013,
  [\href{http://xxx.lanl.gov/abs/1004.1637}{{\tt arXiv:1004.1637}}].

\bibitem{Giannantonio:2009ak}
T.~Giannantonio and C.~Porciani, {\it {Structure formation from non-Gaussian
  initial conditions: multivariate biasing, statistics, and comparison with
  N-body simulations}},  {\em Phys.Rev.} {\bf D81} (2010) 063530,
  [\href{http://xxx.lanl.gov/abs/0911.0017}{{\tt arXiv:0911.0017}}].

\bibitem{Eisenstein:1998hr}
D.~J. Eisenstein, W.~Hu, and M.~Tegmark, {\it {Cosmic complementarity: Joint
  parameter estimation from CMB experiments and redshift surveys}},  {\em
  Astrophys.J.} {\bf 518} (1999) 2--23,
  [\href{http://xxx.lanl.gov/abs/astro-ph/9807130}{{\tt astro-ph/9807130}}].

\bibitem{Schmidt:2010}
F.~Schmidt and M.~Kamionkowski, {\it {Halo clustering with nonlocal
  non-Gaussianity}},  {\em Phys.Rev.} {\bf D82} (2010) 103002,
  [\href{http://xxx.lanl.gov/abs/1008.0638}{{\tt arXiv:1008.0638}}].

\bibitem{Baumann:2012bc}
D.~Baumann, S.~Ferraro, D.~Green, and K.~M. Smith, {\it {Stochastic bias from
  non-Gaussian initial conditions}},  {\em JCAP} {\bf 1305} (2013) 001,
  [\href{http://xxx.lanl.gov/abs/1209.2173}{{\tt arXiv:1209.2173}}].

\end{thebibliography}\endgroup
\bibliographystyle{JHEP}

\end{document}